\UseRawInputEncoding
\documentclass[aps,amsfonts,reprint,tightenlines,amssymb,superscriptaddress,twocolumn, prapplied]{revtex4-2}  
\usepackage{graphicx}
\usepackage{dcolumn}
\usepackage{bm}
\usepackage{mathtools}
\usepackage{stmaryrd}
\usepackage{subcaption}
\usepackage{qcircuit}
\usepackage{caption}
\usepackage[normalem]{ulem}
\usepackage{hhline}

\usepackage{algorithm}
\usepackage[noend]{algpseudocode}
\algrenewcommand\algorithmicdo{}

\makeatletter
\renewcommand{\ALG@name}{Procedure}
\makeatother

\usepackage[linktocpage=true,
  colorlinks=true, 
  pdfborder={0 0 0},
  linkcolor=blue,
  citecolor=red,
  filecolor=yellow,
  urlcolor=blue,
  bookmarks,
  pdfauthor={},
]{hyperref}

\usepackage{orcidlink}
\usepackage{xcolor}
\usepackage{pagecolor}
\usepackage{tabularx}
\usepackage{colortbl} 
\usepackage{xcolor}

\usepackage{ifthen}
\newcounter{is_qcircuit_used}
\begin{document}

\preprint{APS/123-QED}

\title{Demonstration of logical quantum phase estimation for  X-ray absorption spectra}

\author{Hirofumi Nishi\orcidlink{0000-0001-5155-6605}}
\email{hnishi@quemix.com}
\affiliation{
Quemix Inc.,
Taiyo Life Nihombashi Building,
2-11-2,
Nihombashi Chuo-ku, 
Tokyo 103-0027,
Japan
}
\affiliation{
Department of Physics,
The University of Tokyo,
Tokyo 113-0033,
Japan
}

\author{Taichi Kosugi\orcidlink{0000-0003-3379-3361}}
\affiliation{
Quemix Inc.,
Taiyo Life Nihombashi Building,
2-11-2,
Nihombashi Chuo-ku, 
Tokyo 103-0027,
Japan
}
\affiliation{
Department of Physics,
The University of Tokyo,
Tokyo 113-0033,
Japan
}

\author{Satoshi Hirose\orcidlink{0000-0002-6576-9721}}
\affiliation{
Innovative Research Excellence, 
Honda R\&D Co., LTD. \\
Shimotakanezawa 4630, Haga-machi, Haga-gun, 
Tochigi 321-3393, Japan
}

\author{Tatsuya Okayama}
\affiliation{
Innovative Research Excellence, 
Honda R\&D Co., LTD. \\
Shimotakanezawa 4630, Haga-machi, Haga-gun, 
Tochigi 321-3393, Japan
}

\author{Yu-ichiro Matsushita\orcidlink{0000-0002-9254-5918}}
\affiliation{
Department of Physics,
The University of Tokyo,
Tokyo 113-0033,
Japan
}
\affiliation{
Quemix Inc.,
Taiyo Life Nihombashi Building,
2-11-2,
Nihombashi Chuo-ku, 
Tokyo 103-0027,
Japan
}
\affiliation{Quantum Materials and Applications Research Center,
National Institutes for Quantum Science and Technology (QST),
2-12-1 Ookayama, Meguro-ku, Tokyo 152-8550, Japan
}

\date{\today}

\begin{abstract}
In this study, we employed Fourier-based quantum phase estimation (QPE) to calculate X-ray absorption spectroscopy (XAS) spectra. The primary focus of this study is the calculation of the XAS spectra of transition metal $L_{2,3}$-edges, which are dominated by strong correlation effects. First, the Fe $L_{2,3}$-edge X-ray absorption near-edge structure of FePO$_4$ is calculated using a noiseless simulator. The present computation involves a comparison of three types of input states: a uniform superposition state, optimal entangled input state, and Slater function state. Subsequently, we investigated the resolution error of the QPE and statistical error attributed to the measurements. It was revealed that post-processing to introduce Lorentzian broadening reduces the statistical error, which becomes a significant problem for a large number of qubits. Finally, we implemented QPE on a trapped-ion quantum computer, encompassing three orbitals within the active space. To this end, we implemented QPE using dynamic circuits to reduce ancilla qubits and $\llbracket k+2, k, 2\rrbracket$ quantum error detection code to mitigate the quantum noise inherent in current quantum computers. As a result, it was demonstrated that hardware noise was reduced, and spectra close to the noiseless ones were obtained.
\end{abstract}

\maketitle

\section{INTRODUCTION}
X-ray absorption spectroscopy (XAS) is a powerful experimental technique capable of elucidating the local electronic structure and atomic coordination environment of materials on the atomic scale. XAS is extensively used in the study of functional materials such as catalysts \cite{Koningsberger2000TC, Motani2020SAE}, battery materials \cite{Terada2001JSSC, Ghigna2021JPE}, and ferroelectrics \cite{Stern1996JPCS, Yoshida2021APL}, which play an indispensable role in understanding the relationship between structure and function. Accurate theoretical calculations are essential for extracting microscopic information from experimental spectra and validating experimental results.

Density functional theory (DFT) \cite{Hohenberg1964, Kohn1965, Taillefumier2002PRB, Gao2008PRB, Gougoussis2009PRB} and multiple-scattering theory \cite{Joly1999PRL, Joly2001PRB, Joly2009JPCS, Rehr2010PCCP} have traditionally been the main approaches employed for the theoretical analysis of X-ray absorption fine structures (XAFS), including X-ray absorption near-edge structure (XANES) and  extended X-ray absorption fine structure (EXAFS). Although these methodologies successfully reproduce the $K$-edge XANES, they exhibit limitations in accurately capturing many-body interactions and strong electron correlation effects, particularly in the transition metal $L_{2, 3}$-edges dominated by the 2p $\to$ 3d electronic transitions. Recently, several {\it ab initio} wavefunction methodologies have been proposed for the accurate description of $L_{2, 3}$-edge XANES \cite{Kumagai2008PRB, Ikeno2009JPCM, Mizoguchi2010Micron, Ida2012JPCL, Roemelt2013JCP, Ikeno2015MT, Maganas2019JCP}, but such approaches may face difficulties when the number of orbitals is large. Consequently, achieving quantitative agreement with the experimental results often remains challenging.

Recent advances in quantum computing technologies have significantly broadened their potential applications \cite{Nakamura1999Nature, Arute2019Nature, Ming2021Science, Moses2023PRX, Chew2022NatPhot, Bluvstein2024Nature}. Quantum algorithms for high-accuracy quantum chemistry computations on quantum computers demonstrate promising capabilities for accurately incorporating electron correlation effects, which have traditionally posed significant challenges. There are many proposals for quantum algorithms for the ground state \cite{Abrams1997PRL, Peruzzo2014Ncom, Kosugi2022PRR, Nishiya2024PRA, Lin2020Quantum, Dong2022PRXQ} and excited state calculations \cite{Kosugi2020PRA, Kosugi2020PRRes, Cai2020PRR, Huang2022JPCL, Sakuma2024PRA, Fomichev2024arXiv, Fomichev2025arXiv} for an accurate description.

Several types of quantum algorithms have been developed to calculate X-ray absorption spectra (XAS). The first class of algorithms directly evaluates the spectral function in the frequency domain, in which the spectral intensity at each frequency point is obtained using the Hadamard test \cite{Cai2020PRR}. The second approach reconstructs the XAS spectrum using a discrete-time Fourier transform of the measured time-domain response function \cite{Fomichev2024arXiv}. Recently, a highly optimized version of this time-domain algorithm was proposed, achieving the most resource-efficient gate complexity on average \cite{Fomichev2025arXiv}. The third algorithm is based on the quantum phase estimation (QPE) framework, hereafter referred to as QPE sampling \cite{Kosugi2020PRA, Kosugi2020PRRes, Sakuma2024PRA}. Although this approach generally requires a logarithmically greater gate depth than the time domain method, it may reduce the total number of measurement repetitions. However, a detailed quantitative comparison of their computational costs remains to be conducted \cite{Fomichev2024arXiv}.

In the present study, we focus on the implementation of QPE-based quantum algorithms for XAS calculations \cite{Kosugi2020PRA, Kosugi2020PRRes, Sakuma2024PRA} because this approach has the potential to achieve the lowest overall computational cost in the long term.
QPE sampling outputs the transition energies in the ancilla qubits and the probability of obtaining the corresponding energy equal to the intensity of the XAS spectra. Initially, the XAS spectra of Fe $L_{2,3}$-edge XANES of FePO$_4$ were calculated. This material has been extensively studied as a cathode material for lithium-ion batteries. The quality of the XAS spectra obtained by the QPE-sampling is quantitatively evaluated for several variants of QPE \cite{Nielsen2000Book, Luis1996PRA, Fomichev2024arXiv}, including the statistical error arising from measurements. Second, we execute QPE on Quantinuum's trapped-ion quantum computer for the systems using three orbitals in FePO$_4$. To demonstrate the efficacy of the algorithm, we implement QPE using dynamic circuits to reduce the number of ancilla qubits for QPE and the $\llbracket k+2,k,2\rrbracket$ quantum error detection (QED) code, known as the Iceberg code \cite{Steane1996PRA, Gottesman1998PRA, Knill2004arXiv, Knill2004arXiv2, Self2024NPhys, Yamamoto2024PRR, Nishi2025PRApplied}, to mitigate the quantum noise inherent in current quantum devices.

\section{METHOD}
\label{sect:method}
\subsection{A low-energy effective Hamiltonian}
\subsubsection{Spin-Orbit Hamiltonian}
To consider transition from core states, it is necessary to employ an {\it ab initio} Hamiltonian that incorporates relativistic effects.
In this study, we employed the mean-field Breit-Pauli (MFBP) approximation.
We decompose the Hamiltonian into the spin-free (SF) and the spin-orbit coupling (SOC) terms as
\begin{gather}
    \hat{\mathcal{H}}
    =
    \hat{\mathcal{H}}^{\mathrm{SF}}
    +
    \hat{\mathcal{H}}^{\mathrm{MFSO}}
    ,
\end{gather}
where the SF term is
\begin{gather}
\begin{aligned}
    \hat{\mathcal{H}}^{\mathrm{SF}}
    =& 
    \sum_{ij}\sum_{\sigma=\alpha,\beta}
    t_{ij} \hat{a}_{i\sigma}^{\dagger} \hat{a}_{j\sigma}
    \nonumber \\
    & +
    \frac{1}{2} \sum_{ijk\ell}
    \sum_{\sigma, \tau=\alpha,\beta}
    (ij|k\ell) 
    \hat{a}_{i\sigma}^{\dagger}  \hat{a}_{k\tau}^{\dagger}  \hat{a}_{\ell \tau} \hat{a}_{j\sigma}
\end{aligned}
\end{gather}
and the SOC term within the MFBP approximation is given by \cite{Neese2005JCP}
\begin{gather}
    \hat{\mathcal{H}}^{\mathrm{MFSO}}
    =
    \sum_{pq} \boldsymbol{h}^{\mathrm{MFSO}}_{pq}
    \cdot
    \hat{\boldsymbol{T}}_{pq} .
\end{gather}
$\hat{\boldsymbol{T}}_{pq}$ denotes the Cartesian triplet operators and $\boldsymbol{h}^{\mathrm{MFSO}}_{pq}$ is comprised of the one- and two-body SOC, expressed as 
\begin{gather}
    \boldsymbol{h}_{pq}^{\mathrm{MFSO}}
    =
    \boldsymbol{t}_{pq}^{\mathrm{SO}}
    +
    \sum_{rs}
    D_{rs}
    \left(
        \boldsymbol{v}_{pqrs}^{\mathrm{SO}}  
        -
        \frac{3}{2} (
            \boldsymbol{v}_{ps sr}^{\mathrm{SO}}  
            + 
            \boldsymbol{v}_{rqps}^{\mathrm{SO}} 
        )
    \right).
\end{gather}
The definitions of the Cartesian triplet operators and SOC terms are summarized in Appendix~\ref{sect:appendix_hamiltonian}.

\subsubsection{Active space model}
\label{sect:active_space_model}
In this study, the computational protocol comprised two primary steps. Initially, DFT calculations were employed to determine the single-particle electronic structure of the molecular clusters. Subsequently, the automated valence active space (AVAS) method \cite{Sayfutyarova2017JCTC} was utilized to construct an active space. Initially, the molecular cluster FeO$_6^{9-}$ was extracted from the crystal FePO$_4$ with an olivine structure in the Materials Project \cite{MatProj}. 

All quantum chemical calculations were performed using the PySCF package \cite{PySCF2015, PySCF2018, PySCF2020} that employs an atomic orbital basis. To accurately characterize the electronic structure, including the scalar relativistic effect, we utilized the ANO-RCC-VDZP basis set \cite{ANObasis2005} and applied the spin-free exact two-component (sf-X2C) method \cite{X2C1997, X2C2005, X2C2009}. We also employed unrestricted DFT calculations using the hybrid functional with Becke's three parameters and the Lee--Yang--Parr correlation (B3LYP) exchange-correlation functional \cite{B3LYP1988, B3LYP1993, B3LYP1998}.

The Fe $L_{2,3}$-edge XANES is attributed to the transition from the 2p to 3d states. Given the spatial localization of these states, the AVAS method was employed to construct the active space. This method involves projecting predefined atomic valence orbitals onto molecular orbitals obtained from DFT calculations. The AVAS procedure identifies orbitals that exhibit a contribution above a set threshold, thereby ensuring that chemically relevant interactions are comprehensively included. 
The resulting active space was subsequently employed in the XAS spectra calculations using QPE sampling.

\subsection{QPE-sampling}
\label{sect:xas_qpe_sampling}
This study considers an $n_e$-electron system given by the second-quantized many-body Hamiltonian, $\hat{\mathcal{H}}$. 
The ground state and its energy are denoted as $|\Psi_{\mathrm{gs}}\rangle$ and $E_{\mathrm{gs}}$, respectively.
The XAS spectra are expressed as \cite{stefanucci2013Book},
\begin{gather}
    S(\omega)
    =
    -\frac{1}{\pi}
    \mathrm{Im}
    \sum_{\nu =x,z,y}
    \langle \Psi_{\mathrm{gs}}| 
    \hat{\mu}_{\nu}^{\dagger}
    \frac{1}{\omega - \hat{\mathcal{H}} + E_{\mathrm{gs}}  + i\eta}
    \hat{\mu}_{\nu} 
    | \Psi_{\mathrm{gs}}\rangle
    ,
\label{eq:spectral_function}
\end{gather}
where the dipole operator is given by
\begin{gather}
    \hat{\mu}_{\nu}
    =
    \sum_{pq} \mu_{pq}^{\nu}
    \hat{a}_p^{\dagger} \hat{a}_q ,
\label{eq:dipole_opr}
\end{gather}
where $\mu_{pq}^{\nu} = \langle \phi_p | r_{c}^{\nu} | \phi_q \rangle$ and $r_c^{\nu}$ is the position operator.
$\eta$ is a small positive constant denoting the Lorentzian broadening factor.
Orthonormalized one-electron orbitals $| \phi_p \rangle$ in the system are used in this study.
$\hat{a}_p^\dagger$ and $\hat{a}_p$ are the creation and annihilation operators of electrons for these orbitals, respectively.

\begin{figure}[h]
\centering
\mbox{ 
\Qcircuit @C=1em @R=1.2em{
\lstick{|0\rangle}
& \qw & \multigate{2}{U_{\mathrm{in}}} & \ctrl{3} & \qw & \qw & \multigate{2}{\text{QFT}^{\dagger}} & \meter \\
\lstick{|0\rangle}
& \qw & \ghost{U_{\mathrm{in}}}        & \qw & \ctrl{2} & \qw & \ghost{\text{QFT}^{\dagger}}        & \meter \\
\lstick{|0\rangle}
& \qw & \ghost{U_{\mathrm{in}}}        & \qw & \qw & \ctrl{1} & \ghost{\text{QFT}^{\dagger}}        & \meter \\
\lstick{|\Psi_{\mathrm{in}} \rangle} 
& \ustick{\otimes n} \qw   & {/} \qw   &\gate{U} & \gate{U^2} & \gate{U^4} &  \qw                      & \qw \\
}
} 
\caption{
The quantum circuit for the XAS spectra in Eq.~(\ref{eq:spectral_function}) based on the QPE sampling when $n_q=3$. $U=e^{2\pi i t_0 \mathcal{H}/ N_q}$ was used in this figure.
}
\label{circuit:qpe_sampling}
\end{figure}
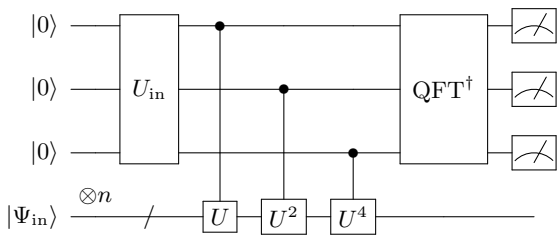

The XAS spectra were calculated on a quantum computer by employing the QPE circuit \cite{Kosugi2020PRA, Kosugi2020PRRes}, as depicted in Fig.~\ref{circuit:qpe_sampling}.
The $n_e$-electron system is encoded to $n$ qubits by employing Jordan-Wigner or Bravyi-Kitaev transformation \cite{Jordan1928, Bravyi2002AnnPhys}. 
We consider the normalized input state on the $n$ qubits given by
\begin{gather}
    |\Psi_{\mathrm{in}}^{\nu}\rangle
    =
    \frac{\hat{\mu}_{\nu} 
    | \Psi_{\mathrm{gs}}\rangle}
    {\|\hat{\mu}_{\nu} 
    | \Psi_{\mathrm{gs}}\rangle\|}
    =
    \sum_{j=0}^{N-1}c_j^{\nu} |\Psi_{j}\rangle ,
\end{gather}
where $N \equiv 2^n$. 
$|\Psi_{\mathrm{in}}^{\nu}\rangle$ is expanded using the eigenvectors of $\hat{\mathcal{H}}$, denoted by $|\Psi_j\rangle$, and $c_j^{\nu}$ denotes the expansion coefficients. By operating the QPE circuit with $n_q$ ancilla qubits, the input state before measurement of the ancilla qubits is expressed as 
\begin{gather}
    \sum_{j=0}^{N-1} c_j^{\nu} 
    \sum_{k=0}^{N_q -1} 
    \alpha\left(
        t_0 (E_j-E_{\mathrm{gs}}) - k
    \right)
    |\Psi_{j}\rangle \otimes |k \rangle_{n_q}
    ,
\label{eq:qpe_output_state}
\end{gather}
where $N_q \equiv 2^{n_q}$ and $|k\rangle_{n_q}$ represent the computational basis of the $n_q$-qubit system. $\alpha$ is a function determined by the adopted unitary $U_{\mathrm{in}}$. $t_0$ represents a scaling parameter that increases or decreases the eigenvalues for precise measurement.
The probability of observing ancilla qubits as $|k\rangle_{n_q}$ state is given by
\begin{gather}
    P_k^{\nu}
    =
    \sum_{j=0}^{N-1} |c_j^{\nu}|^2
    \left|
        \alpha(t_0 (E_j -E_{\mathrm{gs}}) -k)
    \right|^2
\end{gather}

The design of $\alpha (x)$ as a function analogous to the delta function enables loading of the binary representation of each energy eigenvalue into the ancilla qubits of the QPE.
The function $\alpha(x)$ depends on $U_{\mathrm{in}}$, and in conventional QPE,  $U_{\mathrm{in}} = H^{\otimes n_q}$ is used.
Addressing the challenge of spectral leakage (or peak broadening), which arises in conventional QPE with a fewer ancilla qubits, has led to the proposal of entanglement phase estimation (EPE) \cite{Luis1996PRA, Buifmmode1999PRL, van2007PRL, Ji2008IEEE, Babbush2018PRXEncoding, Rendon2022PRD, Sakuma2024PRA} that utilizes the quantum Fourier transform (QFT) as $U_{\mathrm{in}}$. 
Additionally, a recent development in this field involves a method in which the probability distribution of the QPE sampling follows a Lorentzian function by introducing a broadening parameter (QPE-LF) \cite{Fomichev2024arXiv}.
The circuit implementation and behavior of $\alpha(x)$ are detailed in Appendix~\ref{sect:comparison_qpe}.

Given that the absolute squared of the expansion coefficients is equivalent to the transition amplitude, expressed as
$
    |c_i^{\nu}|^{2} 
    = 
    |\langle \Psi_i |\hat{\mu}_{\nu}|\Psi_{\mathrm{gs}} \rangle|^2
    /
    |\langle \Psi_{\mathrm{gs}} |\hat{\mu}_{\nu}|\Psi_{\mathrm{gs}} \rangle|^2
$,  
post-processing using the obtained energies $E_k^{'}$ and the probability $P_k^{\nu}$ reproduces the XAS spectra \cite{Kosugi2020PRA, Kosugi2020PRRes}:
\begin{gather}
    S_{\text{QPE}}(\omega_{\ell}, n_q)
    =
    \frac{1}{C} \sum_{\nu=\{x,y,z\}} \sum_{k=0}^{N_q-1}
    \frac{
        P_k^{\nu}
    }{
        (\omega_{\ell} - E_k^{'} + E_{\mathrm{gs}})^2 
        + 
        \eta^2
    },
\label{eq:qpe_post_process}
\end{gather}
where $C$ denotes the normalization constant such that $\sum_{\ell=0}^{N_{\omega}-1} S_{\text{QPE}}(\omega_{\ell}, n_q) = 1$ for $\omega_{\ell} \in [\omega_{\min}, \omega_{\max})$. In the limit of the large number of ancilla qubits $n_q$, the spectra obtained by QPE are closer to those obtained by the full configuration interaction (FCI)  \cite{stefanucci2013Book}:
\begin{gather}
    S_{\text{FCI}}(\omega_{\ell})
    =
    \frac{1}{\tilde{C}} \sum_{\nu=\{x,y,z\}} \sum_{j=0}^{N-1}
    \frac{|c_j^{\nu}|^2}{(\omega_{\ell} - {E}_j + E_{\mathrm{gs}})^2 + \eta^2},
\end{gather}
where 
$\tilde{C}$ denotes the normalization constant for $S_{\text{FCI}}(\omega_{\ell})$.

We provide a commentary on the preparation of the input state $|\Psi_{\mathrm{in}}\rangle$. Initially, the ground state must be prepared, for which a number of quantum algorithms have been proposed, including the variational quantum eigensolver \cite{Peruzzo2014Ncom}, adiabatic time evolution \cite{Farhi2000arXiv, Nishiya2024PRA}, and probabilistic imaginary-time evolution (PITE) method \cite{Kosugi2022PRR, Nishi2023PRRes, Nishi2024PRR}. 
Subsequent to the preparation of the ground state, the diagonal and nondiagonal parts of the dipole operator are implemented as quantum gates using a linear combination of Pauli operators \cite{Kosugi2020PRA, Kosugi2020PRRes}.
The detail of the implementation of the dipole operator is summarized in Appendix~\ref{sect:appendix_dipole_opr}

\subsection{Hardware implementation}
\subsubsection{Controlled-unitary operation}
In QPE for the $n$-qubit system Hamiltonian $\hat{\mathcal{H}}$, the controlled-real time evolution (RTE) operator is needed to be implemented.
The controlled-RTE for the $(n+1)$ qubits can be implemented as \cite{Yamamoto2024PRR}
\begin{gather}
    \hat{I} \otimes |0\rangle\langle 0| 
    +
    e^{i\theta \hat{\mathcal{H}}} 
    \otimes
    |1\rangle\langle 1|
    =
    \left(
        e^{i\frac{\theta}{2} \hat{\mathcal{H}} \otimes \hat{I}
        }
    \right)
    e^{
        -\frac{i\theta}{2}\hat{\mathcal{H}} \otimes \hat{Z}
    }.
\end{gather}
In this study, we omit the implementation of the bracket term, $e^{i\frac{\theta}{2} \hat{\mathcal{H}}} \otimes \hat{I}$.
We found that the bracket term affects on the output state of QPE in Eq. (\ref{eq:qpe_output_state}) as,
\begin{gather}
    \sum_{j=0}^{N-1} c_j 
    e^{-i\frac{\theta}{2}E_j}
    \sum_{k=0}^{N_q -1} 
    \alpha(t_0 (E_j-E_{\mathrm{gs}}) - k) 
    |\phi_{j}\rangle \otimes |k \rangle_{n_q}
    .
\end{gather}
As is clear from the above equation, the additional RTE does not affect the spectrum obtained by QPE.
In comparison with the implementation of the control RTE operator, this implementation multiplies the coefficient by 0.5, thereby enabling the reduction of the error of the Trotter-Suzuki decomposition by a constant amount. Furthermore, it is noteworthy that the number of CNOT gates remains unchanged in both implementations, irrespective of the implementation method.
This reduction technique was also used in the Hadamard test to estimate ground-state energy \cite{Reiher2017PNAS, Toshio2025PRX}.

\subsubsection{Qubit reduction by $\mathbb{Z}_2$ symmetry}
In this study, we applied a parity transformation \cite{Jacob2012JCP} to the Hamiltonian. This transformation exploits the conservation of the parity in the number of up-spin and down-spin electrons, enabling us to eliminate two qubits from the simulation and thereby recast the original problem into a four-qubit model. Importantly, because the parity transformation explicitly fixes the parity of both spin-up and spin-down electrons, the SOC term, which would otherwise couple these degrees of freedom, is rendered negligible and is subsequently omitted from our calculations. In the following calculation we consider only odd spin-up and spin-down electrons.

To further reduce the computational resources required, we tapered off one qubit by using $\mathbb{Z}_2$ symmetry \cite{Bravyi2017arXiv}. For the four-qubit Hamiltonian $\hat{\mathcal{H}}$, we obtain a symmetry operator $P_{Z_2}=IZIZ$ such that $[\mathcal{H}, P_{Z_2}] = 0$. Using Clifford operator $U$ satisfying $P_{Z_2}=U (IIIX)U^{\dagger}$, the Hamiltonian is transformed as
\begin{gather}
    \hat{\mathcal{H}}^{\prime} 
    =
    U^{\dagger}\hat{\mathcal{H}}U
    =
    \hat{h}_{+} \otimes |+\rangle\langle+|
    +
    \hat{h}_{-} \otimes |-\rangle \langle -|,
\end{gather}
because $[U^{\dagger} \boldsymbol{\sigma}_iU, IIIX] = 0$ for $\hat{\mathcal{H}}=\sum_{i}c_{h_i} \boldsymbol{\sigma}_i$. The eigenvalues are unchanged both before and after the transformation.  For the rightmost qubit, the Hamiltonian has a trivial eigenvalue, which allows for a reduction in the degrees of freedom of the qubit. In XAS calculation, it may occur that an initial and final states belong different eigenspace, e.g., 
$
    |\Psi_{\mathrm{in}}\rangle 
    =
    |\widetilde{\Psi}_{\mathrm{in}}\rangle
    \otimes 
    |+\rangle
$
and
$
    |\Psi_{\mathrm{fin}}\rangle 
    =
    |\widetilde{\Psi}_{\mathrm{fin}}\rangle
    \otimes 
    |-\rangle
$. 
To treat the transitions between different eigenspaces on the QPE-sampling, we decompose the dipole operator as 
\begin{gather}
    \hat{\mu} 
    \to
    \sum_{p,q\in \{+,-\}}
    \hat{\mu}_{pq} \otimes |p\rangle \langle q| .
\end{gather}
Operating the non-diagonal term of the dipole operator leads to
\begin{gather}
    \hat{\mu} |\Psi_{\mathrm{in}}\rangle 
    =
    \hat{\mu}_{-+} |\widetilde{\Psi}_{\mathrm{in}}\rangle \otimes |-\rangle .
\label{eq:input_state_of_qpe_z2_sym}
\end{gather}
The right-hand side of the above equation belongs to the same eigenspace as the final state. Thus, the normalized state in Eq.~(\ref{eq:input_state_of_qpe_z2_sym}) can be used as the input state for the QPE sampling.

\subsubsection{QPE using dynamic circuits}
\begin{figure*}[ht]
    \centering
    \includegraphics[width=0.9 \textwidth]{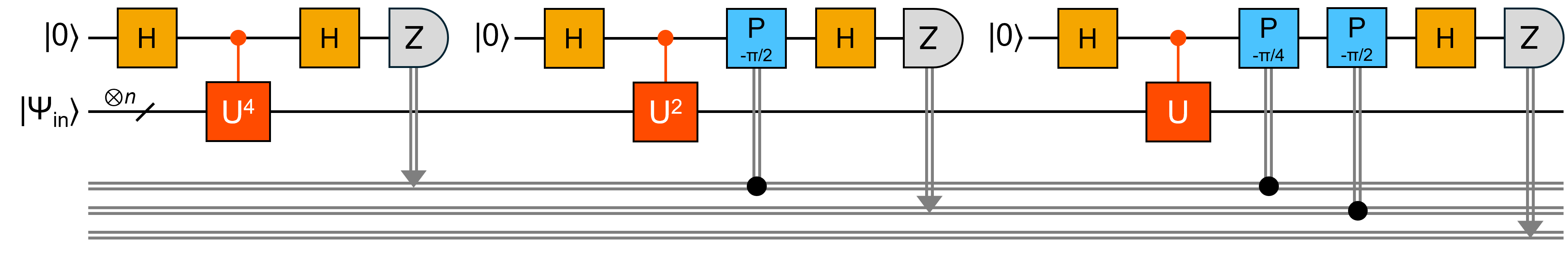} 
\caption{
The quantum circuit for QPE sampling using dynamic circuits corresponding to those for $n_q=3$ and $U_{\mathrm{in}}=H^{\otimes n_q}$ in Fig.~\ref{circuit:qpe_sampling}. The gray double lines indicate the classical registers. Classically controlled phase gates are denoted by $P$ with a rotation angle.
}
\label{fig:qc_dynamic_qpe}
\end{figure*}

In this study, we employed dynamic circuits for QPE, a technique that reduces the number of ancilla qubits and two-qubit gate operations contained in the inverse of the QFT. This reduction is based on the fact that conditional quantum operations just before the measurement can be replaced by classically controlled operations according to the principle of deferred measurement. This dynamic circuit-based implementation is also known as semiclassical QFT \cite{Griffiths1996PRL, Baumer2024PRL} and has been applied to QPE \cite{Higgins2007Nature, Berry2009PRA}. Other ideas related to dynamic circuits are also being applied in other fields such as randomized algorithms \cite{Kanno2025arXiv}. The QPE quantum circuit for QPE is derived using dynamic circuity, as illustrated in Fig.~\ref{fig:qc_dynamic_qpe}. This approach requires a single ancilla qubit,  classically controlled operation, and mid-circuit measurement and reuse.

\subsubsection{Quantum error detection}
The $\llbracket k+2,k,2 \rrbracket$ QED  (Iceberg) code \cite{Self2024NPhys} encodes $k$ even logical qubits $L := \{q_{0}, q_1, \ldots, q_{k-1}\}$ into $k+2$ physical qubits $T := D \cup A$, where $D = L$ and $A := \{q_X, q_Z\}$. The data register $D$ stores the same quantum state as the logical qubit state of $L$, and the Iceberg code employs two redundant qubits $A$ for encoding. A $k$-bit string state $|x\rangle_L = |x_{k-1}\rangle \ldots |x_1\rangle |x_0\rangle$ is encoded as 
\begin{gather}
    |x\rangle_L
    =
    \frac{
        |0 \rangle_{q_Z}
        |f_x \rangle_{q_X} 
        |x\rangle_D
        + 
        |1 \rangle_{q_Z}
        |\neg f_x \rangle_{q_X} 
        |\neg x\rangle_D
    }{\sqrt{2}} ,
\end{gather}
where $\neg x$ is the logical not (negation) of $x$ in the binary representation, $f_x = 0$ for even ${\sum_{i=0}^{k-1}x_i}$, and $f_x =1$ for odd ${\sum_{i=0}^{k-1}x_i}$. The Iceberg code is classified as a stabilizer code, and the code space is stabilized by $S_X = \otimes_{i\in T}X_i$ and $S_Z = \otimes_{i\in T}Z_i$.

In the Iceberg code, the logical Pauli-$X$ and $Z$ gates are defined as
\begin{gather}
\begin{aligned}
    \overline{X}_i
    &=
    X_{q_X} X_i 
    ~~ \forall i \in L 
    \\
    \overline{Z}_i
    &=
    Z_{q_Z} Z_i 
    ~~\forall i \in L ,
\end{aligned}
\end{gather}
where the overline on the operator represents a logical gate. The other logical Puali gates, including the logical Puali-$Y$ gates and logical Pauli gates for multiple qubits, are summarized in Appendix~\ref{sect:appendix_logical_pauli}. The Iceberg code implements logical rotation gates in a transversal manner using a single M\o lmer-S\o rensen (MS) gate \cite{Molmer1999PRL}, defined as
$
    \mathrm{MS}_{ij}(\theta)
    =
    e^{
        -i \frac{\theta}{2}
        Z_i \otimes Z_j
    }
$.
The logical rotation gates $\{\overline{R}_{X}, \overline{R}_{Z}, \overline{R}_{XX}, \overline{R}_{YY}, \overline{R}_{ZZ}\}$ constitute a universal gate set.

\begin{figure*}[ht]
    \centering
    \includegraphics[width=0.85 \textwidth]{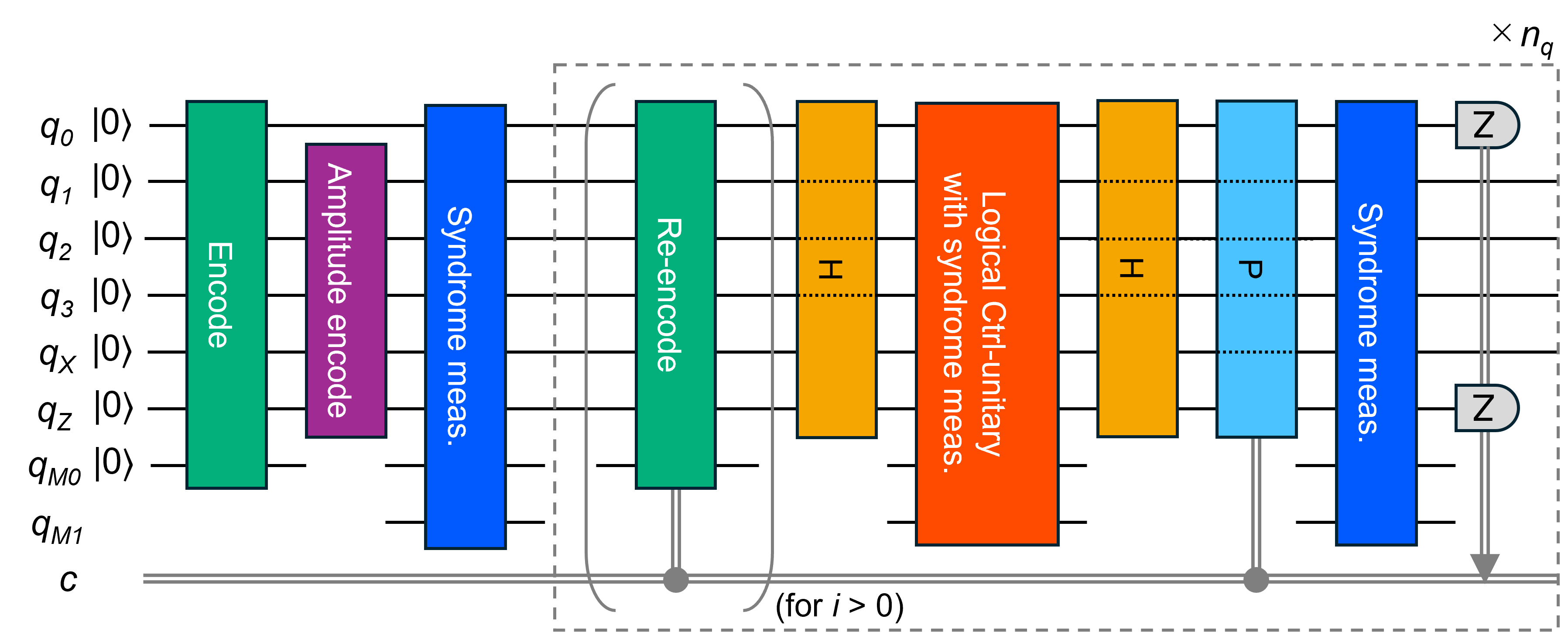} 
\caption{
The quantum circuit for QPE sampling using dynamic circuits illustrated in Fig.~\ref{fig:qc_dynamic_qpe}, which is encoded using iceberg code. Four logical qubits and $n_q=4$ are used in this study.
}
\label{fig:qc_dynamic_qpe_qed}
\end{figure*}

The quantum circuit implemented using the iceberg code is shown in Fig.~\ref{fig:qc_dynamic_qpe_qed}. Eight qubits are used in this study. Initially, the logical zero state was encoded through a circuit that incorporated a flag qubit \cite{Chao2018PRL}. This approach enabled the detection of errors that may arise during the encoding process. Subsequently, we prepared an input state using amplitude encoding. Subsequently, a syndrome measurement was performed, which contained 12 two-qubit gates and two additional qubits. The syndrome measurement of the iceberg code has been engineered to detect errors that occur during syndrome measurement without the need for additional qubits \cite{Self2024NPhys}. Then, we perform logical operations of the QPE using dynamic circuits, where we periodically perform syndrome measurements. In the context of a QPE employing a dynamic circuit, it is imperative to measure the ancilla qubit during circuit execution. However, if this measurement is performed within an encoded circuit, the logical state can be destroyed. A quantum circuit for re-encoding was proposed in \cite{Nishi2025PRApplied}, and this approach is used here.

\section{Results}
\label{sect:simulation}
\subsection{Noiseless simulation results}
\label{sect:results_using_classical}

\subsubsection{Setup}
We applied the QPE sampling algorithm to calculate the XAS spectra of the high-spin state of the FeO$_6^{9-}$ cluster, as described in Sect.~\ref{sect:active_space_model}. The high-spin configuration was selected based on experimental observations \cite{Augustsson2005JCP, Hunt2006PRB, Laffont2006CM}. The active space consists of eight orbitals derived from the Fe 2p and 3d states. These orbitals were selected using the AVAS method, with an overlap threshold of 0.99.

\subsubsection{Results using classical computer}
The electronic configuration of the active space for FeO$_6^{9-}$ is 2p$_{\mathrm{Fe}}^{6}$3d$_{\mathrm{Fe}}^{5}$.
Because the total number of configurations is ${}_{16}\mathrm{C}_{11}=4,368$, we performed CASCI calculations, that is, the FCI within the constructed active space \cite{Roos1980ChemPhys, Zhang2016Book, Norman2018ChemRev}. To examine the accuracy of the QPE for various ancilla input states, the intensity of the XAS spectra was calculated using the CASCI results, that is, ignoring the Trotter-Suzuki decomposition error of the controlled RTE operators. We adopted three types of input states for the QPE ancilla qubits: a uniform superposition state \cite{Nielsen2000Book}, optimal entangled input state \cite{Luis1996PRA, Buifmmode1999PRL, van2007PRL, Ji2008IEEE, Babbush2018PRXEncoding}, and Slater function state \cite{Fomichev2024arXiv}.

To calculate the XAS spectra in the range  $[\omega_{\min}, \omega_{\max})$, we set the scaling factor for the spectra as 
$
    t_0 = N_q / (\omega_{\max} - \omega_{\min})
$.
We also shifted the energy by $\omega_{\min}$ to be plotted from $\omega_{\min}$.
The broadening parameter of the QPE-LF is $a = 2\pi \eta / (\omega_{\max} - \omega_{\min})$ with XAS broadening $\eta$.
In QPE, due to its inherent periodicity, all spectral peaks lying outside the energy window $[\omega_{\min}, \omega_{\max})$ are folded back into this window. In the present study, the peaks outside the interval $[\omega_{\min}, \omega_{\max})$ are negligibly small and thus do not affect the results. However, if significant peaks exist outside this energy range, it becomes necessary to suppress them appropriately e.g., by applying a step function or similar damping mechanism to reduce them to zero \cite{Low2019Quantum, Martyn2021PRXQuantum, Lin2022PRXQ, Kharazi2024arXiv}.
We remark on the procedure used for determining the spectral range. First, the range can be determined by referring to experimental spectra, although this approach relies on prior information. Second, it can be estimated using lower-level theoretical approximations such as 
ROCIS (restricted open shell configuration interaction with singles) \cite{Maganas2013PCCP},
which can provide a rough estimate of the energy region where satellite peaks appear. Third, the method can be applied iteratively, while gradually expanding the spectral range.  On the other hand, employing the core-valence separation (CVS) approximation \cite{Cederbaum1980PRA, Barth1981PRA, Norman2018ChemRev, Michael2020JCP}, which distinguishes core and valence excitations, may not only reduce the gate cost associated with implementing the real-time evolution operator, but also help mitigate aliasing artifacts.

\begin{figure}[ht]
    \centering
    \includegraphics[width=0.45 \textwidth]{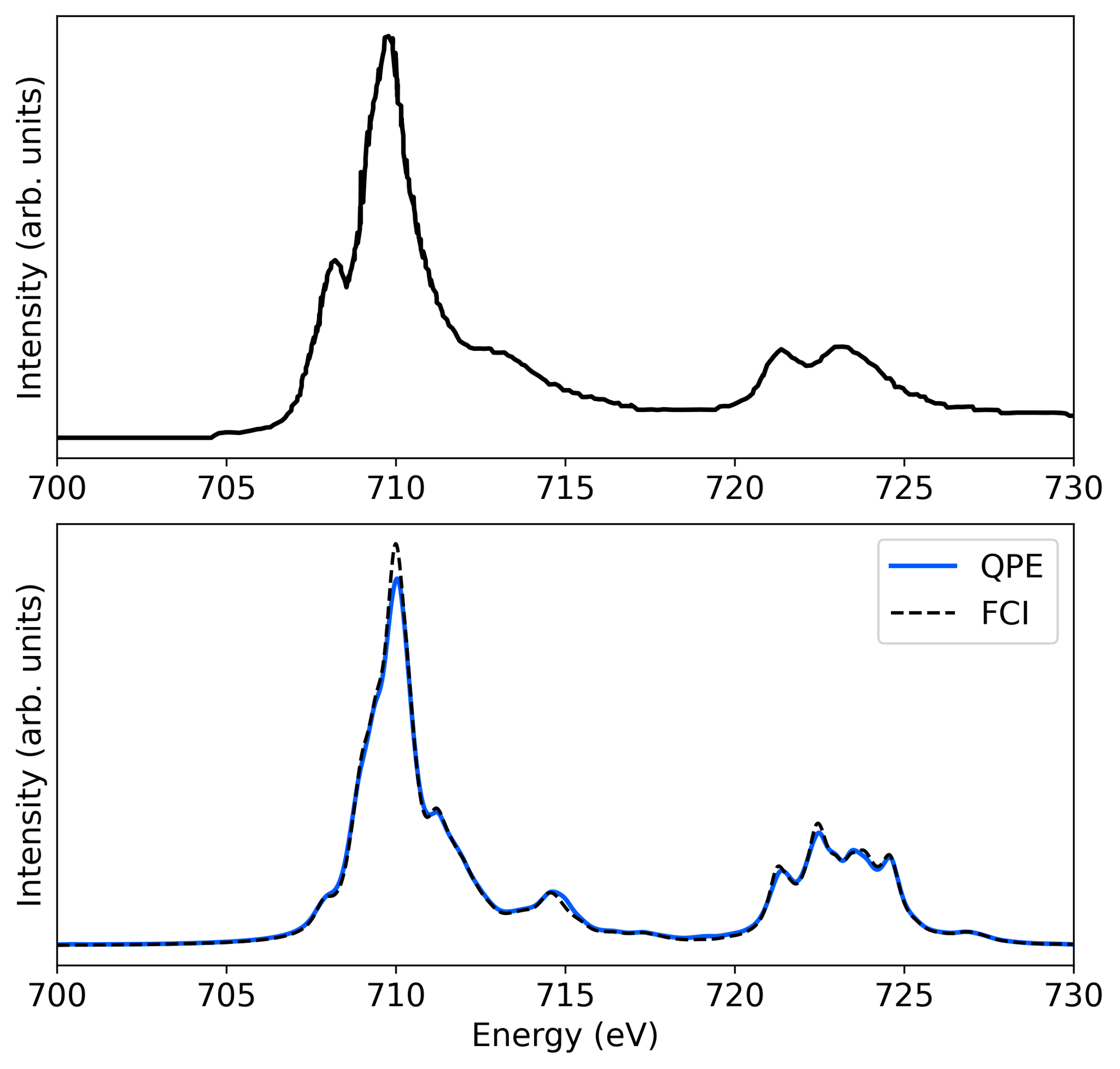} 
\caption{
Calculated Fe-$L_{2,3}$ XANES of FePO$_4$ using CASCI and QPE calculations. We use $\eta=0.3$ eV and $n_q=7$ in this figure. The upper panel shows the experimental spectrum \cite{Augustsson2005JCP}.
}
\label{fig:fci_results}
\end{figure}

The CASCI and QPE results are shown in Fig.~\ref{fig:fci_results}. The spectra from QPE and CASCI are very similar, with only slight differences observed. These arise from an insufficient number of ancilla qubits, and we will quantitatively assess the resulting errors later. Overall, the CASCI and QPE calculations reproduced the experimental spectra well, including the satellite peaks (small peaks around 723 eV) arising from the multiplet term. Here, we introduce a constant energy shift ($-5.04$ eV) to be consistent with the experimental quasi-particle peak. The cause of this constant energy shift was considered to be the small size of the active space. In fact, it has been reported that increasing the active space size of multi-reference {\it ab initio} wavefunction methodologies reduces the energy shift \cite{Maganas2019JCP} and orbital relaxation also reduces the energy shift using $\Delta$SCF or orbital-optimized DFT calculations \cite{Hait2021JPC}. 
Differences between the experimental and calculated spectra were also observed. We consider the differences arising from the interactions of the system outside the FeO$_6$ cluster, the insufficient size of the active space, and/or the exchange-correlation potential.
Additional results obtained using the CASCI calculation are summarized in Appendix \ref{sect:appendix_params_dft}, where we compare the spectra with and without the effective Madelung potential. Also, we show the differences from the exchange-correlation functionals and XAS spectra of LiFePO$_4$.

\begin{figure*}[ht]
    \centering
    \includegraphics[width=1.0 \textwidth]{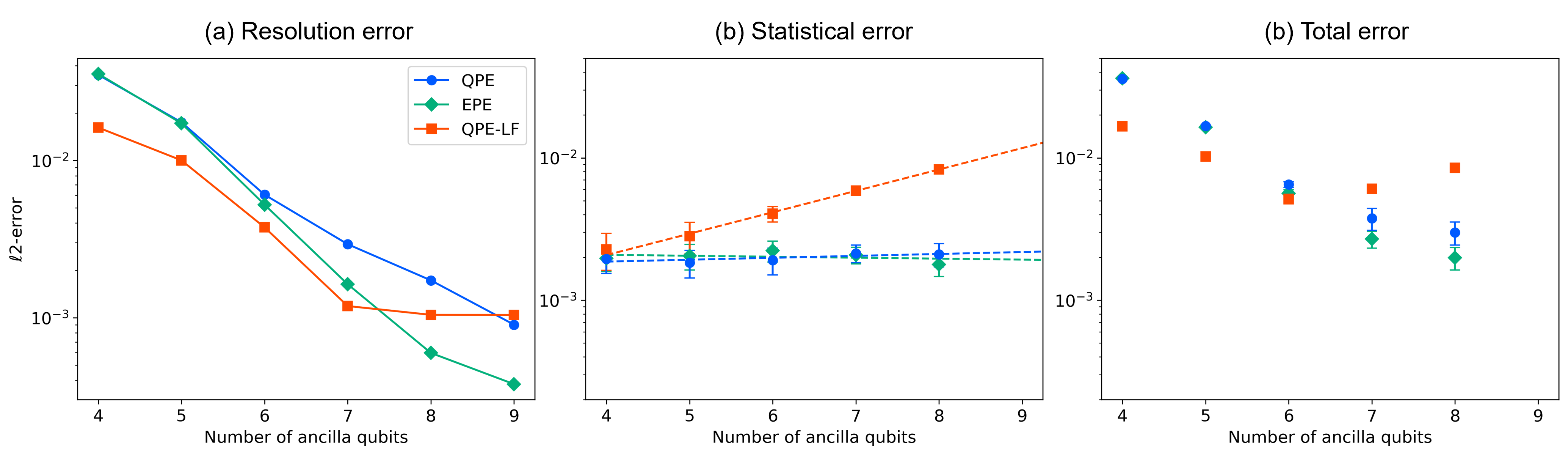} 
\caption{
The $\ell^2$-errors of the three QPE sampling methods with respect to (a) resolution, (b) statistical, and (c) total error according to the number of ancilla qubits. In Figs. (b) and (c), the number of measurements was set to $N_{\mathrm{meas}}=1,000$ and the average of 10 trials was represented as a marker. The standard deviation is indicated by an error bar.
}
\label{fig:qpe_l2_error}
\end{figure*}

We then quantify the accuracy of QPE against the resolution (algorithmic) and statistical error by using $\ell^2$-error defined as
\begin{gather}
    \mathrm{\ell^2error}
    =
    \sqrt{
    \sum_{\ell=0}^{N_{\omega}-1} \left|
        S_{\mathrm{FCI}}(\omega_{\ell})
        -
        S_{\mathrm{QPE}}(\omega_{\ell}, n_{q})
    \right|^2} ,
\end{gather}
where the post-processed spectra $S(\omega_{\ell})$ are adopted. 
We used $\omega_{\ell} = \omega_{\min} + \ell / t_0$ for $\ell =0, 1, \ldots, N_{\omega}-1$ and $N_{\omega}=2^{10}$. We compare three types of QPE: conventional QPE, EPE, and the probability of QPE followed by Lorentz function broadening(QPE-LF). The QPE-LF has already been developed to incorporate Lorentzian broadening; we simply interpolate the resulting spectrum to be consistent with the number of elements in the spectra of $N_{\omega}=2^{10}$.

First, we investigate the dependence of the QPE resolution error on the number of qubits in Fig.~\ref{fig:qpe_l2_error}(a), ignoring the effect of statistical errors due to the measurement. The results demonstrate that the errors in both QPE and EPE exhibit decreasing trends as the number of qubits increases. The findings also reveal that EPE exhibits superior accuracy compared with QPE for all values of qubits. QPE-LF demonstrates optimal accuracy for seven or fewer qubits; however, for higher numbers of qubits, the error converges. This phenomenon can be attributed to the fact that QPE-LF generates a spectrum that closely resembles a Lorentzian distribution, as detailed in Appendix~\ref{sect:comparison_qpe}. The accuracy of the approximation improves as broadening decreases.

Figure~\ref{fig:qpe_l2_error}(b) illustrates the dependence of measurement error on the number of qubits. In this scenario, we compute the $\ell^2$error between the spectrum with a statistical error and that without a measurement error. The results indicate an exponential increase in the $\ell^2$error of the QPE-LF with an increase in the number of qubits. This finding is consistent with the established fact that the larger the number of qubits, the larger the number of measurements required to achieve a fixed accuracy. In contrast, a notable finding is the observation of independence in the effect of statistical errors on the QPE and EPE, irrespective of the number of qubits. This phenomenon is hypothesized to be attributable to the postprocessing step depicted in Eq.~(\ref{eq:qpe_post_process}), which is believed to mitigate statistical errors.

Figure~\ref{fig:qpe_l2_error}(c) shows the $\ell^2$error between the spectrum obtained with a finite number of measurements and the CASCI spectrum. As shown in Fig.~\ref{fig:qpe_l2_error}(a) and (b), in scenarios where the number of qubits is minimal, the predominant factor contributing to the discrepancy is attributed to the resolution error. Conversely, in scenarios where the number of qubits is substantial, the primary contributor to the discrepancy is a statistical error. In addition, the statistical errors shown in Figs.~\ref{fig:qpe_l2_error} (b) and (c) decreased as the number of measurements increases in proportion to $1/\sqrt{N_{\mathrm{meas}}}$.

\subsection{Demonstration on a trapped-ion quantum computer}
\label{sect:demonstration}

\subsubsection{Setup}
Here, we used the same cluster, FeO$_6^{9-}$ in Sect.~\ref{sect:results_using_classical}. However, we consider a low-spin state and only select three orbitals [2p$_z$, 3d$_{z^2}$, 3d$_{xz}$] for execution on a current quantum computer. The electronic configuration of the active space was 2p$_z^{2}$3d$_{z^2}^{0}$3d$_{xz}^{0}$. These orbitals were also selected using the AVAS method, where we set the AVAS overlap threshold to 0.5.

The quantum circuits were constructed using \texttt{qiskit} v1.4.2 \cite{Qiskit}, where the mapping of the fermionic Hamiltonian to spin operators was performed using \texttt{qiskit-nature} v0.7.2. The execution of our circuits was performed on Quantinuum's H1-1 trapped-ion quantum computer utilizing the Azure Quantum service. The Quantinuum H1-1 trapped-ion computer and its emulator (H1-1E) were used in this study \cite{Ryan2022arXiv, Self2024NPhys}. The H1-1 quantum computer is composed of 20 qubits and has the capacity to perform MS gate operations in conjunction with all-to-all connectivity with approximately $\sim 2.0\times 10^{-3}$ infidelity \cite{H1_1}. The state preparation and measurement (SPAM) errors were approximately $3.0\times 10^{-3}$. The quantum circuit with the QED code was submitted using the \texttt{azure-quantum} v3.1.0. Quantum circuits without QED code constructed with \texttt{qiskit} were converted to \texttt{pytket} v2.2.0 \cite{pytket}, after which jobs were submitted using \texttt{pytket-azure} v0.4.0  \cite{AzureQuantum}.

\subsubsection{Results using quantum device}
\label{sect:results_quantum_device}
\begin{figure*}[ht]
    \centering
    \includegraphics[width=1.0 \textwidth]{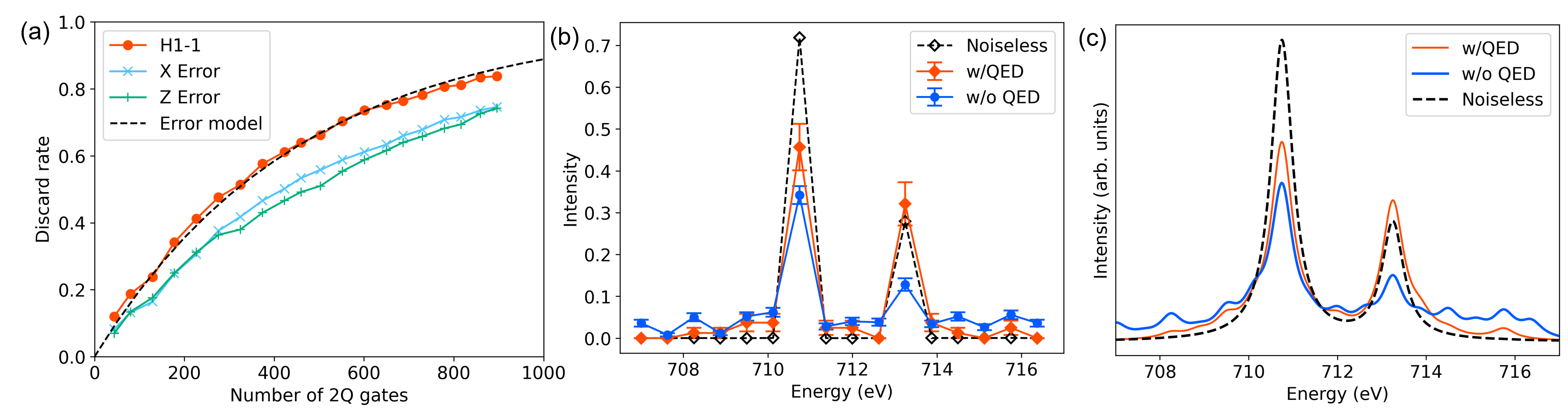} 
\caption{
(a) Discard rates owing to QED-detected errors as a function of the number of two-qubit gates. The markers represent the points at which syndrome measurements were performed. The dashed line denotes the depolarizing noise model, as shown in Eq.~(\ref{eq:depolarizing_model}) with $p_2 = 2.2 \times 10^{-3}$. (b) Raw probability distribution obtained by QPE sampling with and without the QED code. Error bars denote statistical error. (c) Post-processing results for the XAS spectrum obtained using the QPE sampling method shown in (b), where the Lorentzian broadening $\eta=0.3$ eV is used. We ran 500 quantum computational shots.
}
\label{fig:qed_results}
\end{figure*}

The discard rate produced by the QED code was evaluated as a function of the number of two-qubit gates $N_{2Q}$ in Fig.~\ref{fig:qed_results}(a). The underlying hardware noise was modeled with a depolarizing channel, and the empirical fitting form for the discard probability \cite{Yamamoto2024PRR},
\begin{gather}
    d(N_{2Q}, p_2)
    =
    1 - (1 - p_2)^{N_{2Q}} .
\label{eq:depolarizing_model}
\end{gather}
The estimated two-qubit gate error is $p_2 = 2.2 \times10^{-3}$, which is comparable to the value displayed in the hardware specification and obtained in previous studies \cite{Yamamoto2024PRR, Nishi2025PRApplied}. The cumulative detection rates for the Pauli$X$ and $Z$ errors were also plotted. In this experiment, Pauli$X$ errors were detected at a slightly greater frequency. In this experiment, 896 two-qubit gate operations were included and the discard rate was 83.8\%. The XAS spectrum was derived from 81 measurement results in which no errors were detected.

The raw data of the probability distribution obtained by QPE sampling are shown in Fig.~\ref{fig:qed_results}(b). The intensity corresponds to the measurement probability of energy. A first look at the QED shows that the shape of the probability distribution is quite similar to a noiseless shape. In this context, noiseless refers to the exclusion of statistical errors owing to measurements and hardware errors. In the absence of QED, it is evident that the distribution approaches a uniform distribution in its overall shape. The $\ell^2$errors of the spectrum with and without QED were $2.8 \times 10^{-1}$ and $4.3 \times 10^{-1}$, respectively. Although the measurement outcomes that detected errors were discarded in the QED results, thereby enhancing the statistical error, our experiment successfully reproduced the noiseless spectrum.

The results of broadening using the Lorentz function, as given by Eq.~(\ref{eq:qpe_post_process}) are shown in Figs.~\ref{fig:qed_results}(c). In accordance with previously reported findings in Fig.~\ref{fig:qed_results}(b), the spectrum obtained using QED code demonstrates a high degree of similarity to the noiseless spectrum. The $\ell^2$errors of the spectrum with and without QED were $1.8 \times 10^{-2}$ and $2.8 \times 10^{-2}$, respectively. In the subsequent analysis of Sect.~\ref{sect:simulation}, it was confirmed that the statistical error was mitigated by post-processing. However, the effect of mitigating hardware error cannot be clearly observed from the results in Fig.~\ref{fig:qed_results}(c). Consequently, it can be deduced that the implementation of QED is a crucial step in mitigating hardware errors. The results obtained using the emulator are summarized in Appendix \ref{sect:results_of_emulator}. The results obtained in this case are essentially the same as those reported in the main text, but the $\ell^2$errors are smaller than those obtained by H1-1.

\section{CONCLUSIONS}
In this study, we employed QPE to calculate the XAS spectra on a quantum computer. The primary focus of this study is the calculation of the XAS spectra of transition metal $L_{2,3}$-edges, which are dominated by strong correlation effects. First, Fe $L_{2,3}$-edge X-ray absorption near-edge structure of FePO$_4$ were calculated using a noiseless simulator. Subsequently, we investigated the resolution error of the QPE and the statistical error attributed to measurements for three types of input states: a uniform superposition state \cite{Nielsen2000Book}, an optimal entangled input state \cite{Luis1996PRA, Buifmmode1999PRL, van2007PRL, Ji2008IEEE, Babbush2018PRXEncoding}, and a Slater function state \cite{Fomichev2024arXiv}. It was revealed that post-processing to introduce Lorentzian broadening reduces the statistical error, which becomes a significant problem for large qubits. A dependency of the statistical error on the number of qubits was not observed in this calculation.

Next, we implemented QPE on an trapped-ion quantum computer, encompassing three orbitals within the active space. To this end, we implemented QPE using dynamic circuits to reduce ancilla qubits and $\llbracket n+2, n, 2\rrbracket$ QED code to mitigate the quantum noise inherent in current quantum computers. The quantum circuit in this study involves 896 two-qubit gate operations in addition to 52 mid-circuit measurements and reuses.  The QED code was demonstrated to detect quantum errors in 83.8\% of the results. The elimination of erroneous results led to a reduction in the $\ell^2$-error for the post-processed spectrum from $2.8\times 10^{-2}$ to $1.8 \times 10^{-2}$. These calculations are expected to lay the groundwork for material simulations of XAS spectra in the era of fault-tolerant quantum computers.

In the present study, we investigated a mononuclear iron system using an active-space model consisting of eight orbitals on a quantum computer, though the system was already calculated by classical computers \cite{Ikeno2015MT, Wang2023InoChem, Guo2024InoChem} . This size is sufficient for simulations on classical computers; however, current quantum hardware limitations in qubit count and coherence prevent the treatment of larger active spaces. With the anticipated development of quantum hardware, we expect to achieve logical qubit numbers and error-rate levels that will allow significantly larger demonstrations of more complex systems, such as binuclear clusters (e.g., Cu$_2$O) \cite{Levine2020JCTC} and iron-sulfur complexes \cite{Sharma2014NatChem, Mejuto2022JCTC, Lee2023NatComm}. For instance, previous studies \cite{Mejuto2022JCTC} on iron-sulfur clusters (e.g., a [4Fe-4S] cubane with active space of 52 orbitals and 86 electrons) have illustrated the complexity of their electronic structures, suggesting that such systems represent promising targets for future quantum demonstrations.

\section*{DATA AVAILABILITY}
The input crystallographic structures and output spectral data from this study are openly available \cite{Zenodo}.

\section*{ACKNOWLEDGMENTS}
The authors thank Xinchi Huang and Yusuke Nishiya for discussions and valuable comments.
The authors gratefully acknowledge Quantinuum and Microsoft for providing technical support.
In this research work, we used the UTokyo Azure (\url{https://utelecon.adm.u-tokyo.ac.jp/en/research_computing/utokyo_azure/}). 
The authors thank the Supercomputer Center, the Institute for Solid State Physics, the University of Tokyo for the use of the facilities.
This work was partially supported by the Center of Innovations for Sustainable Quantum AI (JST Grant Number JPMJPF2221).

\appendix

\section{Comparison of QPE input}
\label{sect:comparison_qpe}
The QPE based on QFT operates $U_{\mathrm{in}}$ on $n_q$ ancilla qubits before operating the controlled-unitary gates.
The generated state is denoted by
$
    |\psi_{\mathrm{in}}\rangle
    = 
    U_{\mathrm{in}}|0\rangle_{n_q} 
$, where $|k\rangle_{n_q}$ denotes the computational basis for $n_q$ qubits.
After operating the controlled-unitary gates and the inverse QFT, the QPE input state $\sum_{j=0}^{N-1} c_j |\Psi_j\rangle \otimes  |\psi_{\mathrm{in}}\rangle $ is changed to
\begin{gather}
    \sum_{j=0}^{N-1} c_j 
    \sum_{k=0}^{N_q-1} 
    \alpha(t_0 E_j - k) 
    |\Psi_{j}\rangle \otimes |k \rangle_{n_q}
    ,
\label{eq:appendix_psi_qpe}
\end{gather}
where $N \equiv 2^n$ and $N_q \equiv 2^{n_q}$.
$|\Psi_j\rangle$ and $c_j$ represent the $j$th eigenvector of the Hamiltonian $\hat{\mathcal{H}}$ for $n$-qubit system and the expansion coefficients, respectively.
$t_0$ represents a scaling parameter to increase or decrease the eigenvalues for precise measurement.
When $n_q$ ancilla qubits are used to load eigenvalues in the binary representation. 
$\alpha$ depends on the unitary operator $U_{\mathrm{in}}$, and we summarize three choices as follows.

\subsection{A uniform superposition state}
We usually adopted $U_{\mathrm{in}} = H^{\otimes n_q}$ \cite{Nielsen2000Book}, in which the ancilla input state is given by
\begin{gather}
    |\psi_{\mathrm{in}}\rangle
    =
    \frac{1}{\sqrt{N_q}} \sum_{\tau=0}^{N_q-1} |\tau\rangle_{n_q} 
    .
\end{gather}
In this case, $\alpha (x)$ in Eq. (\ref{eq:appendix_psi_qpe}) is calculated as
\begin{gather}
    \alpha (x)
    =
    \frac{1}{N_q} \sum_{\tau=0}^{N_q-1}
    e^{2\pi i \tau x / N_q} .
\label{eq:alpha_qpe}
\end{gather}
We have
\begin{gather}
    |\alpha(x)|^2
    =
    \frac{
        \sin^2 \pi x
    }{
        N_q^2 \sin^2 (\pi x / N_q)
    } .
\end{gather}
When $x=0$, the equation follows $|\alpha(x=0)|^2 = 1$.
However, when $t_0 E_j$ cannot be equal to an integer $k$, the probability has a slight amplitude. 
This phenomena is known as the spectral leakage.

\subsection{Optimal entangled input state}
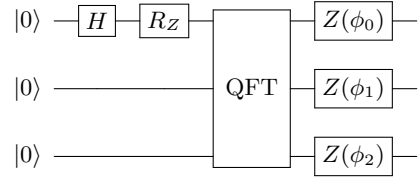
\begin{figure}[h]
\centering
\mbox{ 
\Qcircuit @C=1em @R=1.2em{
\lstick{|0\rangle} & \gate{H} & \gate{R_Z}  & \multigate{2}{\text{QFT}} & \gate{Z({\phi_0})} & \qw \\
\lstick{|0\rangle} &  \qw  & \qw  & \ghost{\text{QFT}} & \gate{Z({\phi_1})} & \qw \\
\lstick{|0\rangle} &  \qw  & \qw  & \ghost{\text{QFT}} & \gate{Z({\phi_2})} & \qw \\
}
} 
\caption{
Quantum circuit for the entangled input state for the EPE in Eq. (\ref{eq:psi_in_epe}) for $n_q = 3$ case. $R_Z = R_Z (\pi/N_q - \pi)$ and $\phi_m = -\pi2^m/N_q$ are used in this figure.
}
\label{circuit:entangled_phase_estimation}
\end{figure}

By tackling the spectral leakage, entangled phase estimation (EPE) is proposed \cite{Luis1996PRA, Buifmmode1999PRL, van2007PRL, Ji2008IEEE, Babbush2018PRXEncoding}. 
The EPE algorithm uses entangled state for the ancilla state, instead of the equal probability distribution state.
The entangled input state is prepared by QFT, given by
\begin{gather}
    |\psi_{\mathrm{in}}^{\mathrm{EPF}} \rangle
    =
    \sqrt{\frac{2}{N_q}} \sum_{\tau=0}^{N_q-1} 
    \sin \frac{\pi (\tau + 1/2)}{N_q} |\tau\rangle_{n_q}  .
\label{eq:psi_in_epe}
\end{gather}
The entangle input state $|\psi_{\mathrm{in}}^{\mathrm{EPF}} \rangle$ is prepared by quantum circuit with QFT \cite{Rendon2022PRD}, depicted in Fig.~\ref{circuit:entangled_phase_estimation}.
The derivation is follows:
\begin{gather}
\begin{aligned}
    R_Z& \left(
        \frac{\pi}{N_q} - \pi
    \right) 
    H |0\rangle_{n_q}
    =
    i
    \frac{
        e^{-i \pi /2N_q}|0\rangle_{n_q} 
        - 
        e^{i \pi /2N_q}|1 \rangle_{n_q}
    }{\sqrt{2}}
    \\ &\overset{\mathrm{QFT}}{\to}
    \frac{i}{\sqrt{2N_q}}
    \sum_{\tau=0}^{N_q-1}
    \left[
        e^{-i \pi /2N_q}
        - 
        e^{2 i \pi (\tau + 1/4) /N_q}
    \right]
    |\tau \rangle_{n_q}
    \\ &=
    \frac{i}{\sqrt{2N_q}}
    \sum_{\tau=0}^{N_q-1}
    e^{i \pi \tau / N_q}
    \left[
        e^{-i \pi \frac{\tau + 1/2}{N_q}}
        - 
        e^{i \pi \frac{\tau + 1/2}{N_q}}
    \right]
    |\tau \rangle_{n_q}
    \\ &=
    \sqrt{\frac{2}{N_q}}
    \sum_{\tau=0}^{N_q - 1}
    e^{i \pi \tau / N_q}
    \sin\left(
        \frac{\pi (\tau+1/2)}{N_q} 
    \right)
    |\tau \rangle_{n_q}
    \\ & \overset{Z_{\phi}}{\to}
    \sqrt{\frac{2}{N_q}}
    \sum_{\tau=0}^{N_q - 1}
    \sin\left(
        \frac{\pi (\tau +1/2)}{N_q} 
    \right)
    |\tau \rangle_{n_q} .
\end{aligned}
\end{gather}
The last equation is derived by applying the phase gates 
$
    \bigotimes_{m=0}^{n_q - 1} Z(-\pi 2^{m} / N_q)
$, 
with $Z(\phi) \equiv Z_{\phi} \equiv \mathrm{diag}(1, e^{i\phi})$. 
$\alpha(x)$ is calculated as
\begin{gather}
\begin{aligned}
    \alpha^{\mathrm{EPE}}(x) 
    & \equiv 
    \frac{\sqrt{2}}{N_q} 
    \sum_{\tau=0}^{N_q-1} e^{2 \pi i \tau x / N_q} 
    \sin \frac{\pi(\tau+1 / 2)}{N_q} 
    \\& \overset{\delta \equiv 2 \pi x}{=}
    \frac{1}{i N_q \sqrt{2}} 
    \sum_{\tau=0}^{N_q-1} 
    \left(
        e^{i \frac{\pi}{2 N_q}} e^{i \frac{\tau(\delta+\pi)}{N_q}}
        -
        e^{-i \frac{\pi}{2 N_q}} e^{i \frac{\tau(\delta-\pi)}{N_q}}
    \right)
    \\ & =
    \frac{1}{i N_q \sqrt{2}}
    \bigg[
        e^{i \frac{\pi}{2 N_q}} 
        \frac{1-e^{i(\delta+\pi)}}{1-e^{i \frac{\delta+\pi}{ N_q}}}
        -
        e^{-i \frac{\pi}{2 N_q}} 
        \frac{1-e^{i(\delta-\pi)}}{1-e^{i\frac{\delta-\pi}{N_q}}}
    \bigg] 
    \\ & =
    \frac{\left(1+e^{i \delta}\right) e^{-i \delta /(2 N_q)}}{N_q 2 \sqrt{2}}
    \left[
        \frac{1}{\sin \frac{\delta+\pi}{2 N_q}}
        -
        \frac{1}{\sin \frac{\delta-\pi}{2 N_q}}
    \right] 
    \\ & =
    - e^{i \delta(1-1 / N_q) / 2} \frac{\sqrt{2}}{N_q} 
    \cos \left(\frac{\delta}{2}\right) 
    \frac{
        \cos \frac{\delta}{2 N_q} \sin \frac{\pi}{2 N_q}
    }
    {
        \sin \frac{\delta+\pi}{2 N_q} \sin \frac{\delta-\pi}{2 N_q}
    } .
\label{eq:alpha_epe}
\end{aligned}  
\end{gather}
The maximum peak hight of $|\alpha^{\mathrm{EPE}}(x=0)|^2$ is lower than one, expressed as
\begin{gather}
    |\alpha^{\mathrm{EPE}}(x=0)|^2
    =
    \frac{2}{N_q^2 \sin^2 \frac{\pi}{2N_q}}
    \to 
    \frac{8}{\pi^2}.
\end{gather}

\subsection{Slater function state}
The spectral function usually adopted the Lorentzian function.
Therefore, Ref.~\cite{Fomichev2024arXiv} proposed the following input state to reproduce the probability of QPE as Lorentzian.
In this case, the input state for ancilla qubits are given by
\begin{gather}
    |\psi^{\mathrm{SF}}_{\mathrm{in}} \rangle
    =
    C_S(n_q, a) \sum_{\tau=0}^{N_q-1} e^{-a\tau}|\tau\rangle_{n_q}  ,
\end{gather}
where the normalization constant 
\begin{gather}
    C_S(n_q, a)
    =
    \sqrt{
        \frac{1 - e^{-2a}}{1 - e^{-2aN_q}}
    } ,
\end{gather}
with positive decay rate $a>0$. In the $a\to 0$ limit,  $|\psi^{\mathrm{SF}}_{\mathrm{in}} \rangle$ becomes the input state of the original QPE. 
The state $|\psi^{\mathrm{SF}}_{\mathrm{in}} \rangle$ is generated by the product of $R_Y$ gates as \cite{Klco2020PRA}
\begin{gather}
    |\psi^{\mathrm{SF}}_{\mathrm{in}} \rangle
    =
    \bigotimes_{m=0}^{n_q - 1} 
    R_Y(2\theta_m) |0\rangle_{n_q}
\end{gather}
where
\begin{gather}
    \theta_m 
    \equiv
    \arctan \exp(-2^m a) .
\end{gather}
Note that the above circuit for the discrete Slater function is a bit different from the original paper \cite{Klco2020PRA}, because the original method is compensated for cusp at the origin. However, here, we only consider positive region $\tau > 0$, so we do not need to consider the cusp.  
The function $\alpha$ is calculated as
\begin{gather}
\begin{aligned}
    \alpha^{\mathrm{LF}} (x)
    & =
    \frac{C_S(n_q, a)}{\sqrt{N_q}}
    \sum_{\tau=0}^{N_q-1}
    e^{-a\tau}
    e^{2\pi i \tau x / N_q}
    \\ &=
    \frac{C_S(n_q, a)}{\sqrt{N_q}}
    \frac{
        1 - e^{2\pi i x - a N_q}
    }{
        1 - e^{2\pi i x / N_q -a }
    }
    \\ & \approx
    \frac{C_S(n_q, a)}{\sqrt{N_q}}
    \frac{1}{a - 2\pi i x / N_q}
\end{aligned}   
\end{gather}
and the absolute squared of $\alpha$ becomes
\begin{gather}
\begin{aligned}
    |\alpha^{\mathrm{LF}} (x)|^2
    &=
    \frac{C_S^2(n_q, a)}{N_q}
    \left|
        \frac{
            1 - e^{2\pi i x - a N_q}
        }{
            1 - e^{2\pi i x / N_q -a }
        }
    \right|^2
    \\ & \approx
    \frac{C_S^2(n_q, a)}{N_q}
    \left|
        \frac{1}{a - 2\pi i x / N_q}
    \right|^2
    \\ & =
    \frac{C_S^2(n_q, a)}{N_q}
    \frac{1}{a^2 + 4\pi^2 x^2 /N_q^2}  .
\label{eq:alpha_slater}
\end{aligned}
\end{gather}
Here we used the following relation:
\begin{gather}
    \left|
        \sum_{\tau=0}^{N_q-1}
        e^{\tau (iy - \eta)}
    \right|^2
    \approx
    \left|
        \frac{1}{y + i\eta}
    \right|^2
    =
    \frac{1}{y^2 + \eta^2} .
\label{eq:appendix_alpha_slater_eq1}
\end{gather}
By introducing $\eta \equiv a T / 2\pi$, Eq.~(\ref{eq:appendix_alpha_slater_eq1}) follows
\begin{gather}
    |\alpha^{\mathrm{LF}}(x)|^2
    =
    \frac{C_S^2(n_q, a) N_q}{(2\pi)^2}
    \frac{1}{x^2 + \eta^2}
\end{gather}
As in the case of the EPE, the limit of $x\to 0$ becomes
\begin{gather}
        |\alpha^{\mathrm{LF}} (x=0)|^2
        =
        \frac{C_S^2(n_q, a)}{N_q a^2} .
\end{gather}
So, the maximum hight is decreased as the decay rate $a$ increases.
Also, in limit of $a \to 0$, $|\alpha^{\mathrm{LF}}(x)|^2$ approaches to $|\alpha(x)|^2$.
For large $N_q$, $C_S^2$ scales as
\begin{gather}
\begin{aligned}
    C_S^2\left(
    n_q, \frac{2\pi\eta}{N_q}
    \right)
    &=
    \frac{
        1 - e^{-4\pi \eta / N_q }
    }{
        (1 + e^{-4\pi \eta / N_q })
        (1 - e^{-2\pi \eta})
    }
    \\ &\to
    O\left(\frac{\eta}{N_q}\right).
\end{aligned}
\end{gather}

\begin{figure}[ht]
    \centering
    \includegraphics[width=0.45 \textwidth]{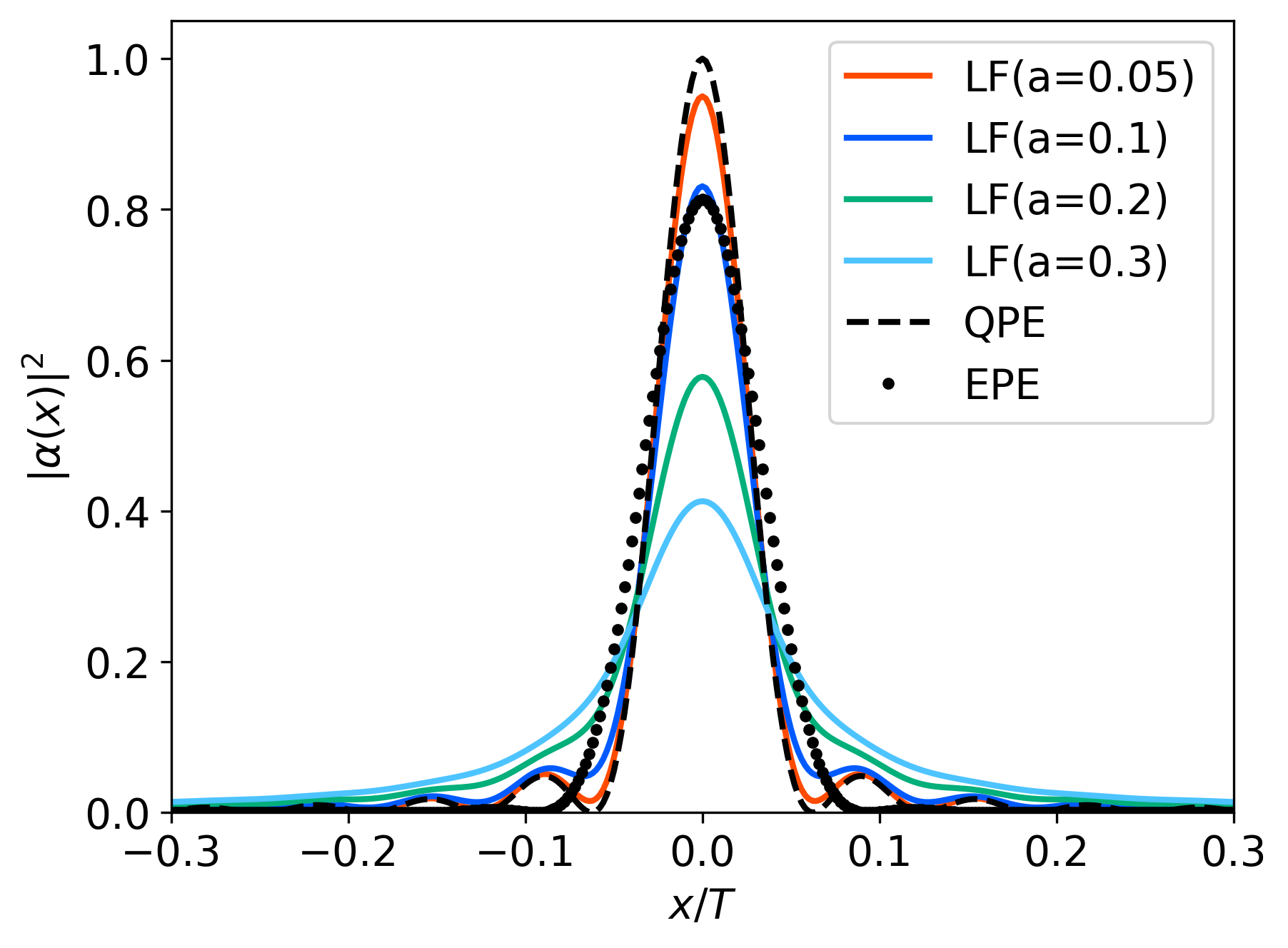} 
\caption{
Comparison of the input state of QPE. Eqs.~(\ref{eq:alpha_qpe}) and (\ref{eq:alpha_epe}) are denoted as QPE and EPE, respectively. Eq.~(\ref{eq:alpha_slater}) is denoted by LF and plotted with several decay rates, $a = 0.05, 0.1, 0.2, 0.3$.  
}
\label{fig:qpe_compare}
\end{figure}

For three kind of $U_{\mathrm{in}}$, we plotted $|\alpha(x)|^2$ in Fig.~\ref{fig:qpe_compare}, where $n_q = 4$.
For original QPE, $|\alpha(x)|^2$ takes maximum value at $x=0$ and rapidly decreases to zero while oscillating at $x \neq 0$. 
Thus, we recognize the spectral leakage for small $n_q$, however this phenomena is suppressed by increasing $n_q$. 
Additionally, we observed that the EPE successfully mitigate the spectral leakage. 
When we adopted the discrete SF state as the ancilla input state, $|\alpha^{\mathrm{LF}}(x)|^2$ exhibits broadening of the spectra.
The broadening is enhanced as $a$ increases.
Also, we observed that $|\alpha^{\mathrm{LF}}(x)|^2$ approaches to the $|\alpha(x)|^2$ of the original QPE and indicates the spectral leakage.
However, for large $n_q$ and adequate large $a$, the spectral leakage is negligible.

\section{Dipole operator implementation}
\label{sect:appendix_dipole_opr}
\subsection{Decomposition of dipole operator}
In this section, we briefly explain the computational cost and implementation of dipole operator $\hat{\mu}$.
The dipole operator has two indices, except for the direction $\nu$, as shown in Eq.~(\ref{eq:dipole_opr}). Thus, four indices appear in the XAS spectra. 
If we calculate the XAS spectra based on each term of the dipole operator by implementing $\hat{a}_p^{\dagger}\hat{a}_q$ as proposed in \cite{Kosugi2020PRRes}, the number of repetitions for all terms of the dipole operator scales as $O(n^4)$ and classical post-processing reproduces the full XAS spectrum.
We aim to reduce the scaling of $O(n^4)$ to implement the dipole operator.

We rewrite the dipole operator as
\begin{gather}
\begin{aligned}
    \hat{\mu}_{\nu}
    &=
    \sum_{ij} \sum_{\sigma,\sigma^{\prime}}
    \mu_{ij}^{\nu} 
    \hat{a}_{i\sigma}^{\dagger} 
    \hat{a}_{j\sigma^{\prime}}
    \\ &= 
    \sum_{i} \mu_{ii}^{\nu}
    +
    \frac{i}{2}\sum_{ij}\sum_{\sigma,\sigma^{\prime}}
    \mu_{ij}^{\nu} \hat{\gamma}_{i,\sigma,0} \hat{\gamma}_{j,\sigma^{\prime},1},
\end{aligned}
\end{gather}
where we introduced the Majorana operator defined as
\begin{gather}
    \hat{\gamma}_{i,\sigma,0} 
    =
    \hat{a}_{i,\sigma} + \hat{a}_{i,\sigma}^{\dagger}
    ,    ~~
    \hat{\gamma}_{i,\sigma,1} 
    =
    -i(\hat{a}_{i,\sigma} - \hat{a}_{i,\sigma}^{\dagger}).
\end{gather}
The Majorana operators satisfies anti-commutation relation as
$
    \{
        \hat{\gamma}_{i,\sigma,x},
        \hat{\gamma}_{j,\sigma^{\prime},y}
    \}
    =
    2\delta_{i,j} \delta_{\sigma,\sigma^{\prime}} \delta_{x,y} \hat{I}_{2}
$
, as in the case of anti-commutator relation of the fermion given by
$
    \{\hat{a}_i, \hat{a}_{j}^{\dagger} \} 
    = 
    \delta_{i,j},
    ~
    \{\hat{a}_i^{\dagger}, \hat{a}_{j}^{\dagger} \}
    =
    \{\hat{a}_i, \hat{a}_{j} \}
    =
    0
$.
Using diagonalization as $\mu^{\nu} = \sum_{k} \lambda_{\nu k} \boldsymbol{u}_{\nu k} \cdot \boldsymbol{u}_{\nu k}^{\mathsf{T}}$, we rewrite the dipole operator as 
\begin{gather}
    \hat{\mu}^{\nu}
    =
    E_{\mathrm{const}}^{\nu}
    +
    \frac{i}{2}\sum_{k} \lambda_{\nu k} 
    \sum_{\sigma,\sigma^{\prime}}
    \hat{\gamma}_{\boldsymbol{u}_{\nu k},\sigma,0} 
    \hat{\gamma}_{\boldsymbol{u}_{\nu k},\sigma^{\prime},1},
\label{eq:dipole_opr_majorana}
\end{gather}
where $E_{\mathrm{const}}^{\nu} \equiv \sum_{i}\mu_{ii}^{\nu}$ and
\begin{gather}
    \hat{\gamma}_{\boldsymbol{u}_{\nu k},\sigma,x}
    \equiv
    \sum_{j=0}^{n-1}
    u_{\nu kj} \hat{\gamma}_{j, \sigma, x} .
\end{gather}
The basis-transformed Majorana operator 
$
    \hat{\gamma}_{\boldsymbol{u}_{\nu k},\sigma,0} 
    \hat{\gamma}_{\boldsymbol{u}_{\nu k},\sigma^{\prime},1}
$ in Eq.~(\ref{eq:dipole_opr_majorana}) is unitary, and can be implemented as described in the next subsection.

\subsection{Majorana operator circuit}
According to Lemma 8 in supplementary materials of Ref. \cite{Von2021PRRes}, the basis-transformed Majorana operator is synthesized as
\begin{gather}
    \hat{\gamma}_{\boldsymbol{u}, \sigma, x}
    =
    U_{\boldsymbol{u}, \sigma, x}^{\dagger}
    \hat{\gamma}_{0, \sigma, x}
    U_{\boldsymbol{u}, \sigma, x}
\end{gather}
with the sequence 
\begin{gather}
    U_{\boldsymbol{u}, \sigma, x}
    =
    \prod_{p=0}^{n-2}
    V_{\boldsymbol{u}, \sigma, x}^{(p)} .
\end{gather}
The each terms of the sequence is given by
\begin{gather}
    V_{\boldsymbol{u}, \sigma, x}^{(p)}
    \equiv
    \exp\left(
        \theta_p[\boldsymbol{u}] \hat{\gamma}_{p, \sigma, x}
        \hat{\gamma}_{p+1, \sigma, x}
    \right),
\label{eq:rotation_majorana}
\end{gather}
where the rotation angle $\theta_p[\boldsymbol{u}]$ is determined by recursively solving the linear chain of equations expressed as
\begin{gather}
\begin{aligned}
    u_0 &= \cos(2\theta_0), \\
    u_1 &= \sin(2\theta_0) \cos(2\theta_1), \\
    &\vdots \\
    u_p &= \cos(2\theta_p) \prod_{j<p} \sin(2\theta_j).
\end{aligned}
\end{gather}

Now, we consider the synthesis of the product of the Majorana operators $
    \hat{\gamma}_{\boldsymbol{u}_k,\sigma,0} 
    \hat{\gamma}_{\boldsymbol{u}_k,\sigma^{\prime},1}
$.
The products $
    \hat{\gamma}_{p,\sigma,x} 
    \hat{\gamma}_{p+1,\sigma,x}
$ and 
$
    \hat{\gamma}_{p,\sigma^{\prime},x} 
    \hat{\gamma}_{p+1,\sigma^{\prime},x}
$ commute with each other for all $\sigma \neq \sigma^{\prime}$ or $x\neq y$. Thus the element of the sequence $V_{\boldsymbol{u}, \sigma, 0}^{(p)}$ satisfies
\begin{gather}
    [
        V_{\boldsymbol{u}, \sigma, 0}^{(p)}, 
        V_{\boldsymbol{u}, \sigma^{\prime}, 1}^{(p)}
    ]
    =
    0.
\end{gather}
Finally, the product is expressed as
\begin{gather}
    \hat{\gamma}_{\boldsymbol{u},\sigma,0}
    \hat{\gamma}_{\boldsymbol{u},\sigma^{\prime},1}
    =
    \left(
        U_{\boldsymbol{u}, \sigma, 0}^{\dagger}
        \hat{\gamma}_{0, \sigma, 0}
        U_{\boldsymbol{u}, \sigma, 0}
    \right)
    \left(
        U_{\boldsymbol{u}, \sigma^{\prime}, 1}^{\dagger}
        \hat{\gamma}_{0, \sigma^{\prime}, 1}
        U_{\boldsymbol{u}, \sigma^{\prime}, 1}
    \right)
    \nonumber \\
    =
    \left(
        \prod_{p=0}^{n-2}
        V_{\boldsymbol{u}, \sigma, 0}^{(p)}
        V_{\boldsymbol{u}, \sigma^{\prime}, 1}^{(p)}
    \right)^{\dagger}
    \hat{\gamma}_{0,\sigma,0}
    \hat{\gamma}_{0,\sigma^{\prime},1}
    \left(
        \prod_{p=0}^{n-2}
        V_{\boldsymbol{u}, \sigma, 0}^{(p)}
        V_{\boldsymbol{u}, \sigma^{\prime}, 1}^{(p)}
    \right)
    \nonumber \\
    =
    \left(
        U_{\boldsymbol{u}, \sigma, 0}
        U_{\boldsymbol{u}, \sigma^{\prime}, 1}
    \right)^{\dagger}
    \hat{\gamma}_{0,\sigma,0}
    \hat{\gamma}_{0,\sigma^{\prime},1}
    \left(
        U_{\boldsymbol{u}, \sigma, 0}
        U_{\boldsymbol{u}, \sigma^{\prime}, 1}
    \right)
\end{gather}
Using Eq.~(\ref{eq:rotation_majorana}) and Jordan-Wigner transformation, the product is implemented as
\begin{gather}
\begin{aligned}
    V_{\boldsymbol{u}, \sigma, 0}^{(p)}
    V_{\boldsymbol{u}, \sigma^{\prime}, 1}^{(p)}
    & =
    e^{
        \theta_p [\boldsymbol{u}]
        (
            \hat{\gamma}_{p,\sigma,0}
            \hat{\gamma}_{p+1,\sigma,0}
            +
            \hat{\gamma}_{p,\sigma^{\prime},1}
            \hat{\gamma}_{p+1,\sigma^{\prime},1}
        )
    }
    \\ & =
    e^{
        -i\theta_p [\boldsymbol{u}]
        (
            \hat{X}_{p,\sigma} \hat{Y}_{p+1, \sigma} 
            -
            \hat{Y}_{p,\sigma^{\prime}} \hat{X}_{p+1,\sigma^{\prime}}
        )
    }.
\end{aligned}
\end{gather}
The basis rotation 
$
    U_{\boldsymbol{u}, \sigma, 0}
    U_{\boldsymbol{u}, \sigma^{\prime}, 1}
$ is implemented as the $O(n)$ sequence of two-qubit rotation gates, as depicted in Fig~\ref{circuit:majorana_rotation}.

\begin{figure}[h]
\centering
\mbox{ 
\Qcircuit @C=1em @R=1.2em{
\lstick{} & \qw & \multigate{1}{R_{xy}(2\theta_0)} & \qw                   & \qw & \qw \\
\lstick{} & \qw & \ghost{R_{xy}(2\theta_0)}        & \multigate{1}{R_{xy}(2\theta_1)} & \qw & \qw \\
\lstick{} & \qw & \qw                   & \ghost{R_{xy}(2\theta_1)}        & \multigate{1}{R_{xy}(2\theta_2)} & \qw \\
\lstick{} & \qw & \qw                   & \qw                   & \ghost{R_{xy}(2\theta_2)}        & \qw \\
\lstick{} & \qw & \multigate{1}{R_{yx}(-2\theta_0)} & \qw                   & \qw & \qw \\
\lstick{} & \qw & \ghost{R_{yx}(-2\theta_0)}        & \multigate{1}{R_{yx}(-2\theta_1)} & \qw & \qw \\
\lstick{} & \qw & \qw                   & \ghost{R_{yx}(-2\theta_1)}        & \multigate{1}{R_{yx}(-2\theta_2)} & \qw \\
\lstick{} & \qw & \qw                   & \qw                   & \ghost{R_{yx}(-2\theta_2)}        & \qw \\
}
} 
\caption{
Quantum circuit for the basis rotation $U_{\boldsymbol{u}, \sigma, 0} U_{\boldsymbol{u}, \sigma^{\prime}, 1}$ for $n_e = 4$ and $(\sigma, \sigma^{\prime}) = (\alpha, \beta)$. The upper half of the qubits is assigned to $\alpha$ spins, while the lower half is assigned to $\beta$ spins.}
\label{circuit:majorana_rotation}
\end{figure}

Since we have
$
    \hat{\gamma}_{0,\sigma,0}
    \hat{\gamma}_{0,\sigma,1}
    =
    iZ_{0, \sigma}
$, we achieve the synthesis of $
    \hat{\gamma}_{\boldsymbol{u},\sigma,0} 
    \hat{\gamma}_{\boldsymbol{u},\sigma,1}
$ with $O(n)$ two-qubit rotation gates. 
For $\sigma \neq \sigma^{\prime}$ case, we implement $\hat{\gamma}_{0,\sigma,0}$ and $\hat{\gamma}_{0,\sigma,1}$ using a linear combination of Pauli operators, as proposed in \cite{Kosugi2020PRA}. The scaling of two-qubit rotation gates is also $O(n)$.

Here we have three possibility to implement the dipole operator. First, we implement the Fermion operator $\hat{a}^{\dagger}_p \hat{a}_{q}$ with a constant depth overhead and $O(n^4)$ repetitions of quantum computation \cite{Kosugi2020PRRes}. 
Second, we implement the Majorana operator 
$
    \hat{\gamma}_{\boldsymbol{u}_k,\sigma,0} 
    \hat{\gamma}_{\boldsymbol{u}_k,\sigma,1}
$ with $O(n)$ depth overhead and $O(n^2)$ quantum computations. 
Third, we directly implemented the dipole operator $\hat{\mu}_{\nu}$ as a linear combination (LCU) of the Majorana operators and avoided repetitions for each term of the dipole operator.
The success probability for the LCU is proportional to the inverse of the number of terms of the LCU as $O(1/n)$; however, the circuit depth only increases as $O(n \log n)$.
Thus, third way is the most efficient way to implement the dipole operator, thus we adopted it.

\section{Spin-orbit Hamiltonian}
\label{sect:appendix_hamiltonian}
We decompose the Hamiltonian into the spin-free and the spin-orbit coupling term as
\begin{gather}
    \hat{\mathcal{H}}
    =
    \hat{\mathcal{H}}^{\mathrm{SF}}
    +
    \hat{\mathcal{H}}^{\mathrm{SO}}
\end{gather}
where 
\begin{gather}
\begin{aligned}
    \hat{\mathcal{H}}^{\mathrm{SF}}
    =& 
    \sum_{ij}\sum_{\sigma=\alpha,\beta}
    t_{ij} \hat{a}_{i\sigma}^{\dagger} \hat{a}_{j\sigma}
    \nonumber \\
    & +
    \frac{1}{2} \sum_{ijk\ell}
    \sum_{\sigma, \tau=\alpha,\beta}
    (ij|k\ell) 
    \hat{a}_{i\sigma}^{\dagger}  \hat{a}_{k\tau}^{\dagger}  \hat{a}_{\ell \tau} \hat{a}_{j\sigma}
\end{aligned}
\end{gather}
and 
\begin{gather}
\begin{aligned}
    \hat{\mathcal{H}}^{\mathrm{SO}}
    =&
    \sum_{pq} \boldsymbol{t}_{pq}^{\mathrm{SO}}
    \cdot
    \hat{\boldsymbol{T}}_{pq}
    \\ &+
    \sum_{pqrs} \boldsymbol{v}_{pqrs}^{\mathrm{SO}} 
    \cdot
    \left(    
        2 \hat{T}_{pr}^{I} \hat{\boldsymbol{T}}_{qs}
        +
        4 \hat{\boldsymbol{T}}_{pr} \hat{T}_{qs}^{I}
    \right) .
\end{aligned}
\end{gather}
$\hat{\boldsymbol{T}}_{ij} = (\hat{T}_{ij}^{X}, \hat{T}_{ij}^{Y}, \hat{T}_{ij}^{Z})$ denote the Cartesian triplet operators given by
\begin{gather}
\begin{aligned}
    \hat{T}_{ij}^{I}
    & =
    \frac{1}{2} \left(
        \hat{a}_{i\alpha}^{\dagger}
        \hat{a}_{j\alpha}
        +
        \hat{a}_{i\beta}^{\dagger}
        \hat{a}_{j\beta}
    \right)
    , \\
    \hat{T}_{ij}^{X}
    & =
    \frac{1}{2} \left(
        \hat{a}_{i\alpha}^{\dagger}
        \hat{a}_{j\beta}
        +
        \hat{a}_{i\beta}^{\dagger}
        \hat{a}_{j\alpha}
    \right)
    ,  \\
    \hat{T}_{ij}^{Y}
    & =
    \frac{1}{2i} \left(
        \hat{a}_{i\alpha}^{\dagger}
        \hat{a}_{j\beta}
        -
        \hat{a}_{i\beta}^{\dagger}
        \hat{a}_{j\alpha}
    \right)    
    , \\
    \hat{T}_{ij}^{Z}
    & =
    \frac{1}{2} \left(
        \hat{a}_{i\alpha}^{\dagger}
        \hat{a}_{j\alpha}
        -
        \hat{a}_{i\beta}^{\dagger}
        \hat{a}_{j\beta}
    \right)
\end{aligned}
\end{gather}
where we defined $\hat{T}_{ij}^{I}$ for convenience.
The one- and two-body terms of $\hat{\mathcal{H}}^{\mathrm{SO}}$ are expressed as
\begin{gather}
\begin{aligned}
    \boldsymbol{t}_{pq}^{\mathrm{SO}} 
    &=
    \langle \phi_p | 
    \hat{\boldsymbol{t}}^{\mathrm{SO}}
    | \phi_q \rangle
    \\ & =
    \frac{1}{2c^2}
    \sum_{\nu=1}^{n_{\mathrm{nucl}}}  
    \int d\boldsymbol{r_1}
    \phi_p^{*}(\boldsymbol{r}_1)
    \frac{
        Z_{\nu} \hat{\boldsymbol{I}}_{1\nu}
    }{
        |\boldsymbol{r}_1 - \boldsymbol{R}_{\nu}|^3
    }
    \phi_q (\boldsymbol{r}_1)
\end{aligned}
\end{gather}
and 
\begin{gather}
\begin{aligned}
    \boldsymbol{v}_{pqrs}^{\mathrm{SO}}  
    & =
    \langle\phi_p \phi_q | 
    \hat{\boldsymbol{v}}^{\mathrm{SO}}
    |\phi_r \phi_{s} \rangle
    \\ & =
    -\frac{1}{2c^2} 
    \int d \boldsymbol{r}_1 d \boldsymbol{r}_2
    \frac{
        \phi_{p}^{*}(\boldsymbol{r}_1) \phi_q^{*}(\boldsymbol{r}_2)
        \hat{\boldsymbol{I}}_{12}
        \phi_{r}^{*}(\boldsymbol{r}_1) \phi_{s}^{*}(\boldsymbol{r}_2)
    }{
        |\boldsymbol{r}_1 - \boldsymbol{r}_2|^3
    }
\end{aligned}
\end{gather}
where $c$ denotes the speed of light, 
$
    \hat{\mathbf{I}}_{i\nu} 
    = 
    (\hat{\boldsymbol{r}}_i - \boldsymbol{R}_{\nu})\times \hat{\boldsymbol{p}}_{i}
$ 
, and 
$
    \hat{\boldsymbol{I}}_{ij} 
    = 
    (\hat{\boldsymbol{r}}_i - \hat{\boldsymbol{r}}_j)\times \hat{\boldsymbol{p}}_{i}
$. 

The mean-field Breit-Pauli approximation \cite{Neese2005JCP} leads to one-body Hamiltonian given by
\begin{gather}
    \hat{\mathcal{H}}^{\mathrm{MFSO}}
    =
    \sum_{pq} \boldsymbol{h}^{\mathrm{MFSO}}_{pq}
    \cdot
    \hat{\boldsymbol{T}}_{pq},
\end{gather}
where
\begin{gather}
    \boldsymbol{h}_{pq}^{\mathrm{MFSO}}
    =
    \boldsymbol{t}_{pq}^{\mathrm{SO}}
    +
    \sum_{rs}
    D_{rs}
    \left(
        \boldsymbol{v}_{pqrs}^{\mathrm{SO}}  
        -
        \frac{3}{2} (
            \boldsymbol{v}_{ps sr}^{\mathrm{SO}}  
            + 
            \boldsymbol{v}_{rqps}^{\mathrm{SO}} 
        )
    \right).
\end{gather}
$D_{rs}$ represents the density matrix and the summation for $r, s$ runs over the occupied orbitals.

\section{DFT Calculation Parameters}
\label{sect:appendix_params_dft}

In this section, we detail the choice of parameters used in our DFT calculations. We examined two exchange-correlation functionals, B3LYP and BP86 (Becke'88 exchange \cite{Beck1988PRA} combined with Perdew'86 correlation \cite{Perdew1986PRB}). In calculations employing the BP86 functional, a level-shift of 0.3 Hartree was imposed to aid convergence.
To incorporate the effective Madelung potential, we constructed an Evjen cell of size $(5,5,5)$ and treated all atoms outside the central FeO$_6$ cluster (which was handled quantum-mechanically) as classical point charges \cite{Evjen1932}. In the Evjen method, atoms located at the vertices, edges, and faces of the supercell are assigned fractional charges of 1/8, 1/4, and 1/2, respectively. The SCF energy difference from (4,4,4) cell fell below 0.22 eV for both BP86 and B3LYP functionals. Accordingly, a (5,5,5) supercell was used for all the subsequent calculations. Convergence of the KS energy levels was likewise confirmed at the (5,5,5) cell size.

\begin{figure}[ht]
    \centering
    \includegraphics[width=0.45 \textwidth]{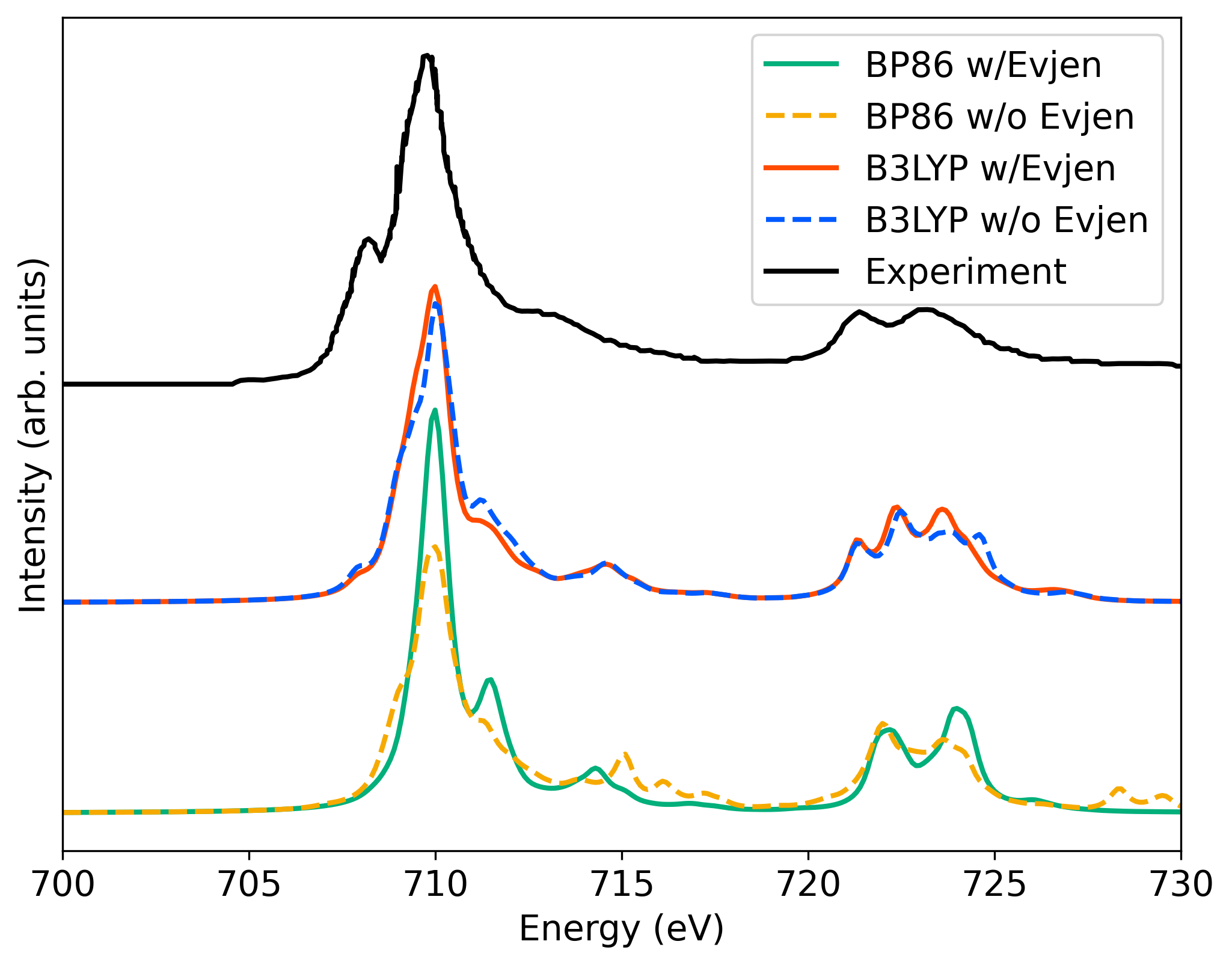} 
\caption{
Experimental \cite{Augustsson2005JCP} and calculated Fe-$L_{2,3}$ XANES of FePO$_4$ using CASCI calculations based on Kohn-Sham orbitals with BP86 and B3LYP functionals. The result compares the spectra with and without the effect of the effective Madelung potential using Evjen's method.
}
\label{fig:xafs_dft_param}
\end{figure}

Figure \ref{fig:xafs_dft_param} compares the computed spectra using both BP86 and B3LYP functionals with and without inclusion of the effective Madelung potential using the Evjen method. 
To be consistent with the experimental quasi-particle peak, we introduce a constant energy shift of -4.96, -10.2 and -11.6 eV for B3LYP functionals, BP86 functional with the Evjen method, and BP86 functional without the Evjen method, respectively.
For the FeO$_6$ cluster with BP86 functional without Evjen treatment, the AVAS method yielded nine active orbitals consisting of the 2p and 3d orbitals; accordingly, we performed CASCI calculations on these nine orbitals.
We observe that upon including the Madelung effective potential, several satellite peak intensities are attenuated relative to the case without a background charge. This trend is consistent with prior studies on LaCoO$_{3}$ and LiCoO$_{2}$ \cite{Kumagai2008PRB}.
Both the BP86 and B3LYP functionals accurately reproduced the XAS spectra. However, slight differences are observed in the satellite peaks.
We consider the differences arising from the interactions of the system outside of the FeO$_6$ cluster, the insufficient size of the active space, and/or the exchange-correlation potential.

\begin{figure}[ht]
    \centering
    \includegraphics[width=0.45 \textwidth]{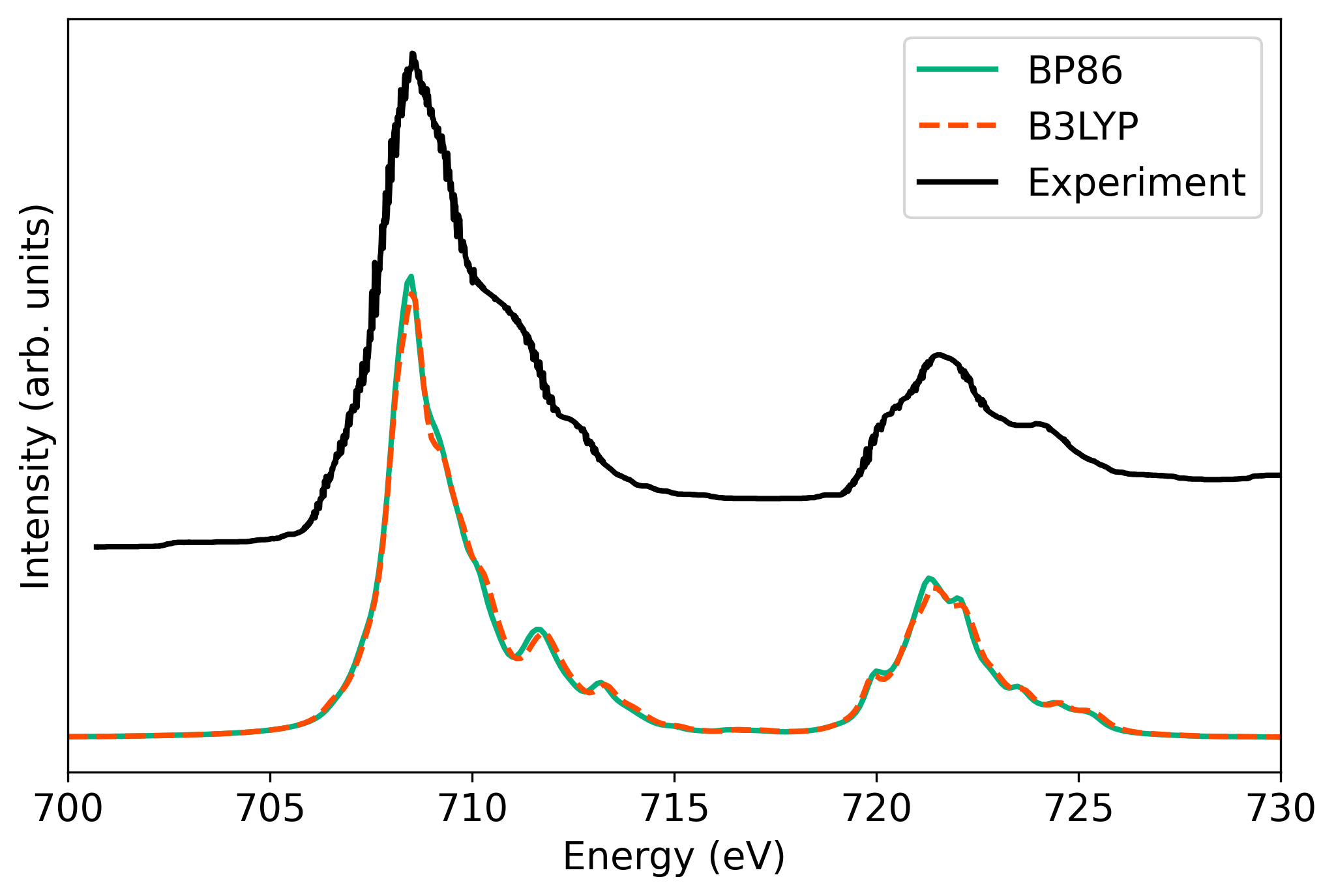} 
\caption{
Calculated Fe-$L_{2,3}$ XANES of LiFePO$_4$ using CASCI calculations with BP86 and B3LYP functionals using the Evjen's method. The experimental spectrum \cite{Augustsson2005JCP} are also plotted.
}
\label{fig:xafs_lifepo4}
\end{figure}

Figure \ref{fig:xafs_lifepo4} shows the XAS spectra of LiFePO$_4$ computed by the CASCI calculation. The FeO$_{6}^{10-}$ cluster was extracted from an olivine-type LiFePO$_4$ crystal structure obtained from the Materials Project \cite{MatProj}. Similar to FePO$_4$, we first performed DFT calculations in the high-spin state, then selected the Fe 2p and 3d orbitals, and constructed an effective Hamiltonian via the AVAS procedure. The CASCI calculations based on this Hamiltonian yielded XAS spectra. Surprisingly, when using BP86 and B3LYP exchange-correlation functionals for the preceding DFT step as reference states, the resulting spectra were nearly indistinguishable. Moreover, both sets of computed spectra agreed well with the experimental data \cite{Augustsson2005JCP}. Therefore, we expect that the QPE sampling can accurately reproduce the experimental XAS spectra of LiFePO$_4$ and FePO$_4$ on a quantum computer.

\section{Logical Pauli gates on the iceberg code}
\label{sect:appendix_logical_pauli}
In the Iceberg code, the logical Pauli-$X$ gates are defined as
\begin{gather}
\begin{aligned}
    \overline{X}_i
    &=
    X_{q_X} X_i 
    ~~ \forall i \in L 
    \\
    \overline{X}_i \overline{X}_j
    &=
    X_i X_j 
    ~~ \forall i, j \in L
    \\
    \otimes_{j \in L \backslash i} \overline{X}_j
    &=
    X_{q_Z} X_i 
    ~~ \forall i L 
    \\
    \otimes_{j \in L} \overline{X}_j
    &=
    X_{q_X} X_{q_Z} ,
\label{eq:logical_pauli_x}
\end{aligned}
\end{gather}
and the logical Pauli-$Z$ gates are
\begin{gather}
\begin{aligned}
    \overline{Z}_i
    &=
    Z_{q_Z} Z_i 
    ~~\forall i \in L
    \\
    \overline{Z}_i \overline{Z}_j
    &=
    Z_i Z_j 
    ~~\forall i, j \in L
    \\
    \otimes_{j \in L \backslash i} \overline{Z}_j
    &=
    Z_{q_X} Z_i 
    ~~\forall i \in L 
    \\
    \otimes_{j \in L} \bar{Z}_j
    &=
    Z_{q_X} Z_{q_Z} .
\label{eq:logical_pauli_z}
\end{aligned}
\end{gather}
Also, the logical Pauli-$Y$ gates are given by
\begin{gather}
\begin{aligned}
    \overline{Y}_i \overline{Y}_j
    &=
    Y_i Y_j 
    ~~\forall i, j \in L 
    \\
    \overline{X}_i \otimes_{j \in L \backslash i} \overline{Z}_j
    &=
    -Y_{q_X} Y_i 
    ~~\forall i, j \in L 
    \\
    \overline{Z}_i \otimes_{j \in L \backslash i} \overline{X}_j
    &=
    -Y_{q_Z} Y_i 
    ~~ \forall i \in L
    \\
    \otimes_{j \in L} \overline{Y}_j
    &=
    (-1)^{1+k / 2} Y_{q_X} Y_{q_Z} .
\label{eq:logical_pauli_y}
\end{aligned}
\end{gather}
Here, the overline on an operator is used to clearly indicate that the operator is a logical gate.

\section{Results of emulator}
\label{sect:results_of_emulator}
\begin{figure*}[ht]
    \centering
    \includegraphics[width=1.0 \textwidth]{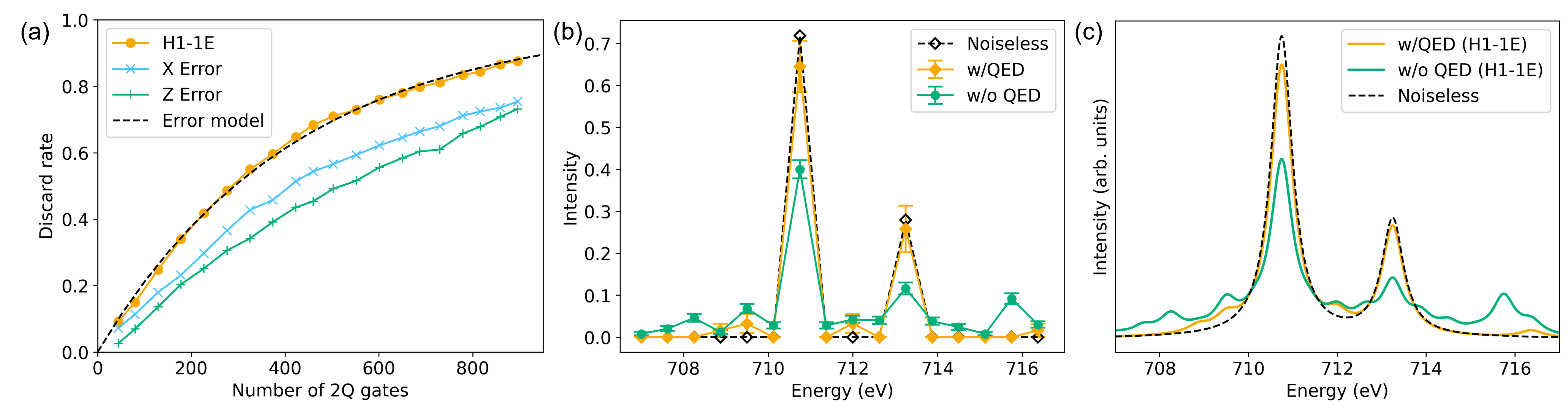} 
\caption{
Same as Fig.~\ref{fig:qed_results} for the emulator of H1-1.
}
\label{fig:result_emu}
\end{figure*}

In this section, we provide the results as the same as Sect.~\ref{sect:results_quantum_device} using the emulator of H1-1.
Figure~\ref{fig:result_emu} represents the discard rate duet to QED code, the probability distribution obtained by QPE-sampling.
We also summarized the results obtained by H1-1 and H1-1E in Table~\ref{tab:results}.

\begin{table}[h]
\centering
\renewcommand{\arraystretch}{1.25}
\caption{This  table summarizes the estimated two-qubits gate error $p_2$, the discard rates $d$ of QED code, the $\ell^2$-error for the probability distribution $P_k$ directly obtained QPE-sampling, and the $\ell^2$-error for the spectra after post-processing.}
\label{tab:results}
\begin{tabular}{lcccc}
\hhline{=====}
 & \multicolumn{2}{c}{H1-1E} & \multicolumn{2}{c}{H1-1} \\
 & w/o QED & w/QED & w/o QED & w/QED \\
\hline
$p_2$ & N/A & $2.4\times 10^{-3}$  & N/A & $2.2\times 10^{-3}$\\
$d$   & N/A & 0.876 & N/A & 0.838 \\
$P_k$'s $\ell^2$-error  & $3.9 \times10^{-1}$ & $9.2\times10^{-2}$ & $4.3\times10^{-1}$ & $2.8\times10^{-1}$
\\
$S(\omega)$'s $\ell^2$-error & $2.6\times 10^{-2}$ & $6.1\times 10^{-3}$ & $2.8\times 10^{-2}$ & $1.8\times 10^{-2}$\\
\hhline{=====}
\end{tabular}
\end{table}

\bibliography{ref}

\end{document}